\documentclass[10pt,a4paper]{article}
\usepackage{qsl-template}
\usepackage{mathrsfs} 
\usepackage[T1]{fontenc} 
\usepackage{lmodern} 
\usepackage{amsmath}
\usepackage{graphicx}
\usepackage{tikz}
\usepackage{etoolbox}
\usepackage{ragged2e}
\usepackage{subcaption}

\usepackage{amsmath, amssymb}
\usepackage{siunitx}
\usepackage[inkscape=false]{svg}
\usepackage{booktabs}
\usepackage{hyperref}
\usepackage{csquotes}
\usepackage[backend=biber,style=numeric-comp,sorting=none, giveninits=true]{biblatex}
\usepackage{comment}

\title{\fontseries{b}\selectfont Systematic Investigation and Suppression of Fluorescence in High-Sensitivity Cavity-Enhanced Raman Gas Sensing}

\AuthorsBlock
{%
\Author{Severin Hager-Roiser}{1\orcid{0009-0007-4821-7443}}\AuthorSep
\Author{Robert Zimmerleiter}{1\orcid{0000-0001-9933-0330}}\AuthorSep
\Author{Paul Gattinger}{1\orcid{0000-0002-0120-9100}}\AuthorSep
\Author{Ivan Zorin}{1\orcid{0000-0002-2089-5716}}\AuthorSep
\Author{Johannes D. Pedarnig}{2\orcid{0000-0002-7842-3922}}\AuthorLastSep
\Author{Markus Brandstetter}{1\orcid{0000-0002-8679-8097}}%
}
{%
\Affil{1}{Research Center for Non-Destructive Testing, Linz, Austria}\\
\Affil{2}{Johannes Kepler University, Institute of Applied Physics, Linz, Austria}%
}
{severin.hager-roiser@recendt.at}

\begin{document}
\maketitle

\abstract{
Raman spectroscopy enables broadband, multi-species gas analysis by providing access to an entire vibrational spectrum in a single measurement. However, the sensitivity of gas-phase Raman sensing is often limited by weak signals and fluorescence background from various optical elements that constrain the achievable signal-to-noise ratio (SNR) through signal-dependent noise contributions (e.g. shot noise). 
Here, we present a cavity-enhanced Raman spectroscopy (CERS) gas sensor employing a \SI{500}{mW}, \SI{532}{nm} continuous wave (CW) laser and a simple, non-resonant two-mirror multi-pass cavity (MPC) operated at ambient pressure and near the concentric condition, providing up to 45 internal reflections. 
To quantitatively capture the impact of fluorescence on performance, a CCD-specific noise model was developed that links fluorescence-induced baseline levels to measurement noise. 
Complementary optical simulations were employed to assess the signal collection efficiency in the MPC.
Through a systematic analysis of fluorescence sources, the background was reduced substantially by step-wise elimination of fluorescent optics. 
The fluorescence-minimized setup resolves weak Raman signatures in ambient-air spectra, including CO$_2$ peaks, O$_2$ and N$_2$ overtones, and ambient CH$_4$ (\rm{$\sim$}\SI{2}{ppm}). Calibration measurements for O$_2$ (diluted in N$_2$), N$_2$ (in O$_2$) and H$_2$ (in N$_2$) demonstrate detection limits of \SI{11}{ppm}, \SI{5}{ppm} and \SI{3}{ppm}, respectively, with a \SI{180}{s} measurement time. The results highlight fluorescence mitigation as a key design lever for robust, field-oriented CERS instrumentation for trace gas sensing.
}

\section{Introduction}
Trace gas analysis has been a highly important research topic for many decades with applications ranging from environmental monitoring~\cite{monks_atmospheric_2009, petrov_multipass_2022},
material science~\cite{jiao_outgassing_2019}, food safety~\cite{lehotay_application_2002}, vacuum systems monitoring~\cite{repa_analyses_2002} to inline process control~\cite{huang_-line_2017}. Especially today, where atmospheric and ambient air quality are topics of increasing concern~\cite{monks_atmospheric_2009} and hydrogen technologies gain significance both in the fuel~\cite{farias_use_2022, rolo_hydrogen-based_2023} and in the industrial
production sector~\cite{souza_filho_green_2022}, fast and reliable gas sensing technologies play a crucial role. \\
Gas chromatography (GC) remains a reference method for complex mixtures down to ultra-low concentrations~\cite{kaminski_determination_2003, bartle_history_2002, van_ruth_methods_2001}, but its cycle time and sampling logistics are often incompatible with fast feedback control and distributed sensing. Electrochemical sensors are highly sensitive but suffer from poor selectivity as a result of cross-sensitivities and are prone to signal drifts~\cite{gebicki_application_2016, wilson_applications_2009}.

Optical spectroscopy offers alternative in-situ measurement techniques capable of high selectivity, sensitivity and fast response times. Infrared (IR) absorption methods are mature, compact, and offer low limits of detection (LOD)~\cite{fu_recent_2022, kreuzer_ultralow_1971, okeefe_cavity_1988}. Yet they are intrinsically insensitive to homonuclear diatomic gases, such as O$_2$, N$_2$ (main components of ambient air) and H$_2$ because their dipole moments remain unchanged during molecular vibrations~\cite{long_raman_2002} and only orders of magnitude weaker quadrupole ro-vibrational transitions add to the absorption signal~\cite{herzberg_possibility_1938, herzberg_quadrupole_1949}. Furthermore, IR-based approaches either require multiple lasers, expensive tunable lasers, or broadband supercontinuum lasers~\cite{zorin_advances_2022} if multiple gases are to be measured simultaneously at low concentrations. Raman spectroscopy, on the other hand, provides access to all gases with only a few exceptions (e.g. noble gases). Raman system architectures require only a single monochromatic laser source, but can quantify multiple components simultaneously from one spectrum at reasonably short measurement times.

For gases, the main limitation of spontaneous Raman spectroscopy is the small scattering cross section~\cite{long_raman_2002, weber_raman_1979, fenner_raman_1973} combined with the low molecular density, which makes the detected signal weak at low concentrations. Several optical enhancement methods have been developed to increase the effective interaction length or collection efficiency. In cavity-enhanced Raman spectroscopy (CERS), the excitation laser is passed through the sample many times in order to increase the number of scattering events. This can be realized in a resonant optical cavity (ROC) with large intracavity power build-up, or in a non-resonant multi-pass cavity (MPC) that increases the path length by repeated reflections without a resonance condition~\cite{wang_review_2020, niklas_short_2021}.
High-finesse ROCs can achieve very large enhancement factors. Yang et al. built a Fabry-Pérot cavity with a finesse of 80000, reaching an intracavity laser power of up to \SI{1000}{W} with an input laser power of \SI{240}{mW}, and obtained impressive detection limits of less than \SI{0.15}{ppm} for N$_2$, O$_2$ and H$_2$~\cite{yang_multiple_2023} with an exposure time of \SI{500}{s}. Wang et al. achieved LOD values less than \SI{30}{ppm} for N$_2$ and O$_2$ and around \SI{5}{ppm} for H$_2$ with an exposure time of \SI{20}{s} with a cavity in a V-shaped configuration using 3 mirrors~\cite{wang_multigas_2020} and an intracavity laser power of \SI{222}{W}.
However, ROCs require active frequency locking or passive optical feedback locking to maintain resonance between laser and cavity~\cite{wang_review_2020, niklas_short_2021, black_introduction_2001, nickerson_review_2019}. The resulting sensitivity to vibration and thermal drift can be problematic for field-oriented measurement systems.
Furthermore, the high-reflectivity cavity mirrors absorb much of the signal light before transmitting it~\cite{yang_multiple_2023}.

Non-resonant multi-pass cavities (MPC) provide simpler, mechanically and thermally more stable alternatives without the need for locking mechanisms. Despite the non-resonant character, impressive signal enhancements have been achieved, especially when the analyte gas is pressurized, which is difficult to realize with ROCs. P. Wang et al. designed a Z-shaped MPC consisting of 4 mirrors with the collection of Raman scattered light collinear with the laser beam. They reached extremely low LODs of \SI{3}{ppb} for CH$_4$ and around \SI{25}{ppb} for N$_2$ and O$_2$ with a \SI{300}{s} measurement time, an analyte pressure of \SI{25}{bar}~\cite{wang_cavity-enhanced_2023} and a laser power of \SI{10}{W}. The same group also used a Herriott-type MPC with 2 mirrors and a pressurization to \SI{5}{bar} to achieve LODs in the single-digit ppm range~\cite{wang_highly_2023} using an exposure time of \SI{200}{s} and a laser power of \SI{1.3}{W}. \\
A very common and simple type of MPC consists of 2 mirrors in near-concentric configuration, where the laser is coupled into the cavity close to the edge of one cavity mirror~\cite{xiao-yun_diagnosis_2008, singh_ambient_2023, singh_high-precision_2023, singh_isotopic_2021, arachchige_raman_2024, velez_spontaneous_2021, zheng_gas_2024, muktha_arachchige_portable_2025, gomez_velez_trace_2020, yang_high-sensitivity_2024, guo_high-sensitivity_2021, yang_highly_2016, petrov_multipass_2016, petrov_multipass_2022, li_near-confocal_2008, miao_parabolic_2024} instead of guiding it through a hole in a mirror (Herriott cell). This approach maintains a low sample gas volume, compared to more complex geometries, while still generating high Raman signals. Singh and Muller employed such a configuration together with a \SI{2.4}{W} blue laser and pressurization to \SI{2}{bar} reaching a detection limit of \SI{50}{ppb} for H$_2$  for a \SI{10}{min} exposure time~\cite{singh_high-precision_2023}.

Another challenge for high-sensitivity Raman spectroscopy is fluorescence. Even when the gas itself is non-fluorescent, many common optical materials and coatings fluoresce under visible excitation and contribute a broadband background in the detected spectrum~\cite{singh_ambient_2023}. This background is detrimental in multiple ways: it reduces spectral contrast, saturates the detector at long integration times, and increases noise, e.g. through shot noise. In a multi-pass geometry with collection collinear with the laser beam, fluorescence from optical components in the beam path can therefore become the dominant noise source, particularly for long integration times required for trace gas detection. A conceptually simple mitigation is to collect Raman light from the side of the cavity to avoid fluorescence originating from the collinear beam path. However, side collection also restricts the effective collection volume and can remove much of the signal advantage of the MPC.

In this work, we address the problems of applied high-sensitivity Raman spectroscopy and propose a tailored multi-faced approach to signal enhancement via optimized optical scheme, signal collection, fluorescence minimization, and detection. An explicit focus is placed on fluorescence minimization and quantitative noise analysis.
Hence, the paper reports on a two-mirror-MPC-based CERS system for high-sensitivity gas analysis (including homonuclear diatomic gases). Thus, we investigate several technical aspects, such as: (i) a detector noise model for the employed CCD spectrometer that links the spectral background level to noise and therefore to achievable detection limits; (ii) a Raman scattering simulation of collinear versus side collection in a two-mirror MPC to assess the viability of side collection as a fluorescence-mitigation strategy; (iii) an experimental demonstration that replacing fluorescent optical components (including cavity mirrors) reduces the fluorescence background in the Raman spectra by about one order of magnitude; and (iv) representative spectra and calibration curves for O$_2$, H$_2$ and N$_2$ with detection limits of \SI{11}{ppm}, \SI{5}{ppm} and \SI{3}{ppm}, respectively, for a \SI{180}{s} measurement time. Due to its simplicity, the sensor can be miniaturized into a portable field-deployable device with minor modifications. Further optimization of low-fluorescence optics and system hardware could achieve sub-ppm detection limits for highly sensitive, real-time gas analysis. 

\section{Materials and Methods}
\subsection{Measurement Concept and Setup}
\begin{figure}[hb]
 \centering
 \includegraphics[width=0.99\textwidth]{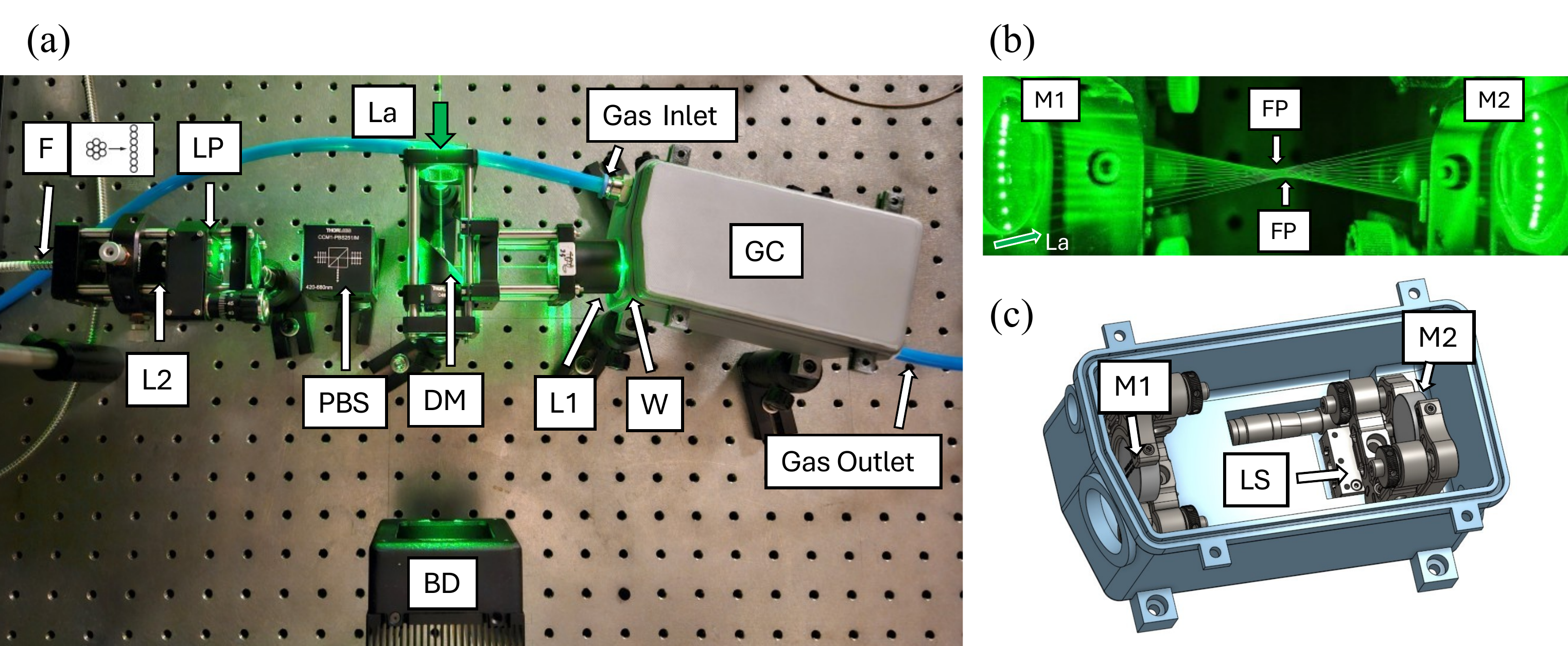}
 \caption{\hyperref[fig:setup]{a} Non-resonant two-mirror MPC CERS setup with collinear signal collection. The excitation beam is launched into the MPC (enclosed in a custom gas cell GC) and performs $N$ reflections before exiting the MPC on its input path. Raman scattered photons are separated from the excitation laser at a dichroic mirror and coupled into a round-to-linear multimode fiber after polarization and long-pass filtering. \hyperref[fig:setup]{b} Typical beam pattern within the MPC. \hyperref[fig:setup]{c} custom designed 3D printed gas cell with included mirror mounts and linear translation stage. Abbreviations: La: laser; DM dichroic mirror; L1: focusing/collimation lens; W: window; M1/M2: cavity mirrors; PBS: polarizing beam splitter; LP: long-pass filter; L2: coupling lens; F: fiber; FP: focal point; GC: gas cell; LS: linear translation stage; BD: beam dump.}
 \label{fig:setup}
\end{figure}

The optical arrangement of the sensor system is shown in \hyperref[fig:setup]{Figure 1a}. The measurement principle is based on spontaneous Raman scattering excited by a linearly polarized narrow-line continuous-wave (CW) laser (Cobolt 05-01 Samba from \textit{HÜBNER Photonics GmbH}) with a wavelength of \SI{532}{nm}, a beam quality of M$^2$$<1.1$, a linewidth of $<$\SI{500}{kHz} and an optical power of \SI{500}{mW}. Enhancement is achieved by repeatedly passing the excitation beam through the sample gas in a custom 3D-printed gas cell (\hyperref[fig:setup]{Figure 1c}). The laser beam is focused by an uncoated UV-enhanced fused silica (UVFS) $f=$ \SI{75}{mm} lens L1 (LA4725 from \textit{Thorlabs}) into the center of a non-resonant two-mirror multi-pass cavity (MPC), which consists of two concave $f=$ \SI{25}{mm} dielectric mirrors (CM254-025-E02 from \textit{Thorlabs}) at near-concentric configuration.\\
The laser beam undergoes $N$ reflections producing a dense pattern of foci (see \hyperref[fig:setup]{Figure 1b}) that increases the effective interaction volume without imposing a resonance condition. After $\frac{N+1}{2}$ reflections, the beam retraces its own path and after $N$ reflections it leaves the MPC exactly on its input trajectory. In combination with the signal collection geometry collinear with the beam path, this has the advantage that both forward and backward scattered Raman light can be detected.\\
Furthermore, collinear collection is well suited for MPC CERS because a significant fraction of the light scattered along the beam path can be re-imaged along the multi-pass trajectory and coupled efficiently into the detection system. The trade-off is that collinear collection also collects fluorescence generated anywhere along the optical beam path. 
The number of reflections $N$ is set by the injection position and angle and is adjusted via mirror alignment. Up to $N=45$ reflections could be realized in this MPC but it was found that an optimal balance between signal chipping losses at the mirror edges and the number of reflections could be achieved for $N=37$, yielding the highest signal.
The Raman signal and the laser beam are collimated by L1 and separated at the dichroic mirror DM (LPD02-532RU from \textit{Semrock}), where the laser beam is reflected back towards the source and the signal beam is transmitted. A polarizing beam splitter (CCM1-PBS251 from \textit{Thorlabs}) suppresses largely unpolarized fluorescence while transmitting the predominantly polarized Raman signal. A long-pass filter (86225 from \textit{Edmund Optics}) provides additional rejection of any remaining laser light. The $f=$ \SI{12.7}{mm} lens L2 (62561 from \textit{Edmund Optics}) couples the signal into a round-to-linear multimode fiber (BFL200RS2 from \textit{Thorlabs}). The fiber is connected to a lens-grating-lens spectrometer (532X -R from \textit{Wasatch Photonics}) with a slit width of \SI{25}{\um} and an f-number of f/1.3. It incorporates a CCD detector with 2048 pixels cooled down to 10°C. The spectrometer has a spectral range of approximately \SI{270}{cm^{-1}} to \SI{4770}{cm^{-1}}, a specified wavenumber resolution of \SI{10}{cm^{-1}} and a 16-bit signal resolution.

\subsection{Signal Scaling and Multi-Pass Enhancement}
For a given Raman transition, the intensity $I$[W/sr] of the generated Raman signal can be written in a compact form as~\cite{long_raman_2002, niklas_short_2021, wang_review_2020}
\begin{equation}
\label{eq: intensity}
I=\mathfrak{I}_0 \sigma'\rho V,
\end{equation}
where $\mathfrak{I}_0$ [W/m$^2$] is the irradiance of the excitation laser, $\sigma' \ \mathrm{[m^2/sr]} \propto \frac{1}{\lambda_s ^4}$ is the differential cross section of a given transition for a certain molecule, $\lambda_s$ is the wavelength of the scattered light, $\rho$ [m$^{-3}$] is the molecular density of the sample molecules and $V$ is the effective scattering volume. \\
If $R_L$ is the reflectance of the MPC mirrors at the laser wavelength, then the laser beam segment after the $i^{th}$ reflection has an irradiance of $\mathfrak{I}_i=\mathfrak{I}_0 R_L^i$. Now we can imagine that each new beam segment contributes an effective scattering volume $\Delta V_{bs}$. We now introduce the abbreviation $a=\sigma' \rho V_{bs}\Delta\Omega_d \eta$ [m$^2$], where $\Delta\Omega_d$ is the solid angle of detectable radiation scattered in the volume $\Delta V_{bs}$, either in forward or in backward direction, and $\eta$ is the efficiency of the collection optics and the spectrometer. Then the signal power $P_i$ [W] detected by the spectrometer for the $i^{th}$ beam segment becomes:
\begin{equation}
    P_i = \mathfrak{I}_0 R_L^i a\left[\, \underbrace{R_s^i}_{\mathrm{backward}} + \underbrace{R_s^{N-i}}_{\mathrm{forward}} \, \right]
\end{equation}
where $R_s$ is the reflectance of the MPC mirrors for the wavelength of the Raman scattered light. The factors $R_s^i$ and $R_s^{N-i}$ arise because the backward scattered light retraces the laser beam path and undergoes $i$ reflections before leaving the MPC and the forward scattered light travels along with the laser beam and experiences $N-i$ reflections before leaving the MPC.\\
Thus, the total collected signal power $P_s$ of the MPC is given by:
\begin{equation}
\label{eq: signal vs refl}
    P_s = \sum_{i=0}^N \mathfrak{I}_0 R_L^i a \left[\, \underbrace{R_s^i}_{\mathrm{backward}} + \underbrace{R_s^{N-i}}_{\mathrm{forward}} \, \right] + P_{\mathrm{ol}} = \mathfrak{I}_0 a \left[ \underbrace{ \frac{1-(R_LR_S)^{N+1}}{1-R_LR_S}}_{\mathrm{backward}} + \underbrace{\frac{R_S^{N+1} - R_L^{N+1}}{R_S - R_L}}_{\mathrm{forward}} \right] + P_{\mathrm{ol}}
\end{equation}
The extra term $P_{\mathrm{ol}}$ accounts for effects related to the overlap of the beam segments in the focal points of the MPC~\cite{arachchige_raman_2024, singh_ambient_2023}, which has been neglected so far. In the last step, the sum formula for geometric series has been used. A similar formula has been derived by Wang et al.~\cite{wang_highly_2023}. For the mirrors used in this work, $R_L\approx0.996$ and $R_s\approx 0.99475$ for the wavelength of the laser and the N$_2$ Raman transition, respectively, according to the data supplied by the manufacturer. Compared to the most basic setup with no cavity mirrors, only one beam segment ($N=0$) and collection of only backward scattered radiation, this formula (using $N=37$ and neglecting $P_{\mathrm{ol}}$) predicts an enhancement factor of around 63. The experimentally observed value, however, is 75, outperforming the prediction despite obvious chipping losses at the mirror edges. The discrepancy is explained by the overlap term $P_{\mathrm{ol}}$ with the following reasoning: 
there are two focal points in the center of the cavity where the individual beam segments overlap (see \hyperref[fig:setup]{Figure 1b}). In the discussion above, each beam segment was assigned an effective scattering volume $\Delta V_{bs}$, neglecting beam overlaps. For the two focal points, it is more comprehensible to consider the laser irradiance. If $\frac{N}{2}$ beam segments pass through each of these points, then the irradiance is locally increased by a factor of $\frac{N}{2}$ for $R_L\rightarrow1$. As the volume of each of the two focal points is shared by $\frac{N}{2}$ beam segments, light scattered in the propagation direction (and opposite direction) of each of the $\frac{N}{2}$ beam segments can reach the detector, yielding another factor of $N$ for $R_S\rightarrow1$. A superlinear enhancement (\text{$\propto N^2$}) therefore results in these points~\cite{arachchige_raman_2024, singh_ambient_2023}. This effect also contributes to the high signal collection from the cavity center in the simulations of \autoref{sec: simulations}.

\subsection{Spectral Processing, Peak Quantification and Limits of Detection}
Unless noted otherwise, spectra are only dark-corrected and averaged to a fixed total acquisition time. For calibration measurements, six spectra with \SI{30}{s} integration time each are averaged, corresponding to \SI{180}{s} total acquisition time. These parameters were chosen using an Allan--Werle-type noise analysis (see Figure S1 in the \hyperref[sec: SI]{Supplementary Information (SI)}): for several integration times (\SI{5}{s} to \SI{60}{s}), the overlapping Allan--Werle deviation reached a minimum at total averaging times of around \SI{100}{s} to \SI{200}{s}, with the best signal-to-noise ratio (SNR) obtained for \SI{30}{s} integration averaged six times (\SI{180}{s}). \\
Raman peaks are quantified by fitting Gaussian line shapes to the vibrational Q-branch features and extracting the peak area. The validity of using Gaussian fits for Raman peaks has previously been confirmed by Wang et al.~\cite{wang_cavity-enhanced_2023}. Calibration curves are obtained by weighted linear regression of Raman peak area versus concentration. Limits of detection (LODs) are reported using two definitions to reflect common practice and to capture imperfections of the calibration. The standard IUPAC-style definition is~\cite{noauthor_iupac_nodate, long_limit_1983}:
\begin{equation}
\label{eq: iupac}
\mathrm{LOD}_{\mathrm{std}} = \frac{k s_b}{b_1},
\end{equation}
where $s_b$ is the standard deviation of blank measurements, $b_1$ is the calibration slope, and the numerical factor $k=3$ corresponds to a 99.7\% confidence interval assuming a Gaussian noise distribution. A corrected definition (LOD$_{\mathrm{pred}}$) accounts for uncertainty of the calibration procedure and possible non-linearities in the data by calculating the 99.7\% prediction bands for the calibration curve~\cite{mermet_calibration_2010, mermet_limit_2008, draper_applied_1998, massart_handbook_1997}. The lowest detectable signal is given by the intersection of the upper prediction band and the intercept (see \autoref{fig: calibratino curves}):
\begin{equation}
\label{eq: LOD_pred}
    \mathrm{LOD}_{\mathrm{pred}} = \frac{t s_{p}}{b_1},
\end{equation}
where $s_{p}$ is the estimated standard deviation of prediction and $t$ is a numerical value for 99.7\% confidence drawn from the Student’s t distribution.

\subsection{Sample Gas Preparation}
Analyte gas mixtures are produced by dynamic dilution using two mass flow controllers (see Figure S2 in the \hyperref[sec: SI]{SI}). A premixed gas from \textit{Linde Gas GmbH} (analyte diluted in N$_2$ or O$_2$, depending on the calibration) is delivered by a low-flow controller and diluted by a high-flow controller supplying high-purity diluent. The target concentration $c$ is calculated from the premix concentration $c_p$ and the premix and diluent flow rates $F_p$ and $F_d$ as
\begin{equation}
c \approx \frac{c_p F_p}{F_p + F_d},
\end{equation}
assuming complete mixing and negligible analyte in the diluent. Concentration is reported in ppm on a volumetric basis. The sample gas mixture is then fed into the custom 3D-printed gas cell (\hyperref[fig:setup]{Figure 1c}). The gas exchange time of the gas cell was characterized by switching from laboratory air to a pure diluent flow and monitoring the N$_2$ Raman signal  (see  Figure S3 in the \hyperref[sec: SI]{SI}). At a maximum flow rate of \SI{1}{l/min} the concentration decayed exponentially with a time constant of about \SI{21}{s} reaching the 1\% level after roughly \SI{110}{s}. Hence, to ensure a complete gas exchange for subsequent calibration measurements, a \SI{4}{min} relaxation time was maintained after each concentration change.

\subsection{Noise Model for CCD-limited Measurements}
To quantify how the fluorescence background affects sensitivity, we employed a CCD noise decomposition adapted from Irie et al.~\cite{irie_technique_2008}. For fixed acquisition settings, such as integration time, the standard deviation $\sigma_N$ of repeated measurements at a given pixel can be modelled as
\begin{equation}
\sigma_N = \sqrt{a \mu_S + b \mu_S^2 + c}, \label{eq: noise model}
\end{equation}
where $\mu_S$ is the mean detected signal in counts, $a \mu_S$ captures shot-noise, $b \mu_S^2$ accounts for photo-response non-uniformity (PRNU) noise, and $c$ aggregates constant noise terms (readout, quantization and fixed pattern noise as well as dark-current contributions for fixed integration time); $a,b,$ and $c$ are noise magnitude coefficients. As the spectrometer does not apply digital filters, corresponding multiplicative noise types are neglected. In this framework, a fluorescence background increases $\mu_S$ even when no Raman features are present and therefore increases $\sigma_N$, directly reducing the achievable SNR and increasing the LOD.

\subsection{Scattering Simulations}
Scattering simulations were performed in the optical design and simulation software \textit{Zemax}. The optical arrangement of the MPC has been recreated in the non-sequential mode in two configurations: in collinear and side detection geometry (\autoref{fig: simulations}). The simulated MPC was set to achieve only 19 laser beam reflections for better visual clarity. The laser beam was represented by a bundle of rays with a Gaussian intensity distribution. Raman scattering is realized via the volume scattering mechanism available in \textit{Zemax}, which scatters light omnidirectionally with a given probability in a certain volume. A small cylindrical object (length: \SI{0.5}{mm}; radius: \SI{0.3}{mm}) served as a scattering volume and was moved across the MPC area to analyze the amount of scattered signal collected from each point within the MPC. 

\section{Results and Discussion}
\subsection{Noise Behavior and the Role of Fluorescence Background}
\label{subsec: noise analysis}
\begin{figure}
     \centering
     \begin{subfigure}[b]{0.49\textwidth}
         \centering
         \caption{}
         \includegraphics[width=\textwidth]{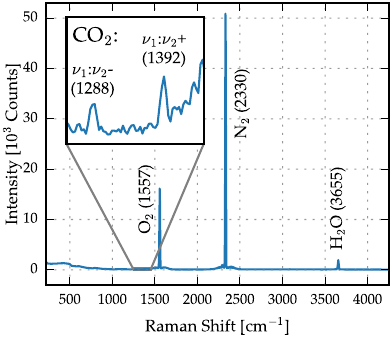}
         \label{subfig: spectrum air 500ms}
     \end{subfigure}
     \hfill
     \begin{subfigure}[b]{0.49\textwidth}
         \centering
         \caption{}
         \includegraphics[width=\textwidth]{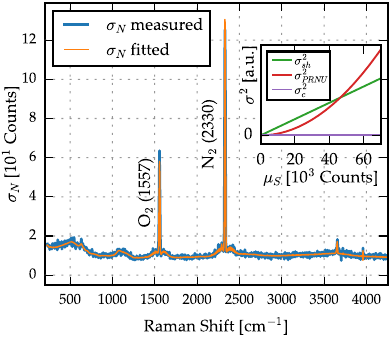}
         \label{subfig: noise spectrum}
     \end{subfigure}
\setlength{\abovecaptionskip}{0pt} 
 \caption{\subref{subfig: spectrum air 500ms} Raman spectrum of laboratory air acquired with \SI{500}{ms} integration time and averaged 10 times (\SI{5}{s} total). The O$_2$, N$_2$ and H$_2$O vibrational Q-branches are indicated and the inset shows the Raman peaks of CO$_2$. \subref{subfig: noise spectrum} Per-pixel standard deviation of 100 Raman spectra of laboratory air with \SI{500}{ms} integration time (blue) and fit using the CCD noise model from \autoref{eq: noise model} (orange). The visibility of Raman features in the noise trace illustrates that background-dependent shot noise and PRNU noise directly impact Raman peak SNR. The inset shows how the individual types of noise scale with the signal mean $\mu_S$.}
 \label{fig:noise}
\end{figure}

\autoref{subfig: spectrum air 500ms} shows a typical spectrum of laboratory air acquired with the setup shown in  \hyperref[fig:setup]{Figure 1a} at an integration time of \SI{500}{ms} averaged 10 times. The O$_2$ and N$_2$ vibrational peaks at \SI{1557}{cm^{-1}} and \SI{2330}{cm^{-1}}, respectively, are clearly visible~\cite{weber_raman_1979, fenner_raman_1973} and even the Fermi diad of CO$_2$, consisting of two peaks at \SI{1288}{cm^{-1}} ($\nu_1:\nu_2-$) and \SI{1392}{cm^{-1}} ($\nu_1:\nu_2+$), is detected~\cite{weber_raman_1979, fenner_raman_1973}.
\autoref{subfig: noise spectrum} serves to display analysis of the measured noise behavior of the spectrometer for the MPC CERS system for 100 individual spectra of laboratory air with \SI{500}{ms} integration time. The measured per-pixel standard deviation $\sigma_N$ (blue) is in good agreement with the CCD model in \autoref{eq: noise model} (orange), with fitted parameters $a=0.19$, $b=4\times10^{-6}$ and $c=26.7$ for the used acquisition settings. The inset in \autoref{subfig: noise spectrum} illustrates how constant noise $\sigma_c$ only has a negligible influence for such short integration times, even at small signals, while shot noise $\sigma_{sh}$ becomes the dominant contribution over the mid-signal range. At sufficiently high signals, PRNU-type noise $\sigma_{PRNU}$ becomes the main noise source. The signal-dependent behavior is highlighted by the fact that the Raman peaks themselves appear in the noise trace. As a consequence, any fluorescence background that raises the baseline also raises the noise floor and significantly decreases the sensitivity, emphasizing the importance of fluorescence minimization.

\subsection{Collinear Versus Side Collection in a Two-Mirror MPC}
\label{sec: simulations}
\begin{figure}
     \centering
     \begin{subfigure}[b]{0.49\textwidth}
         \centering
         \caption{}
         \includegraphics[width=\textwidth]{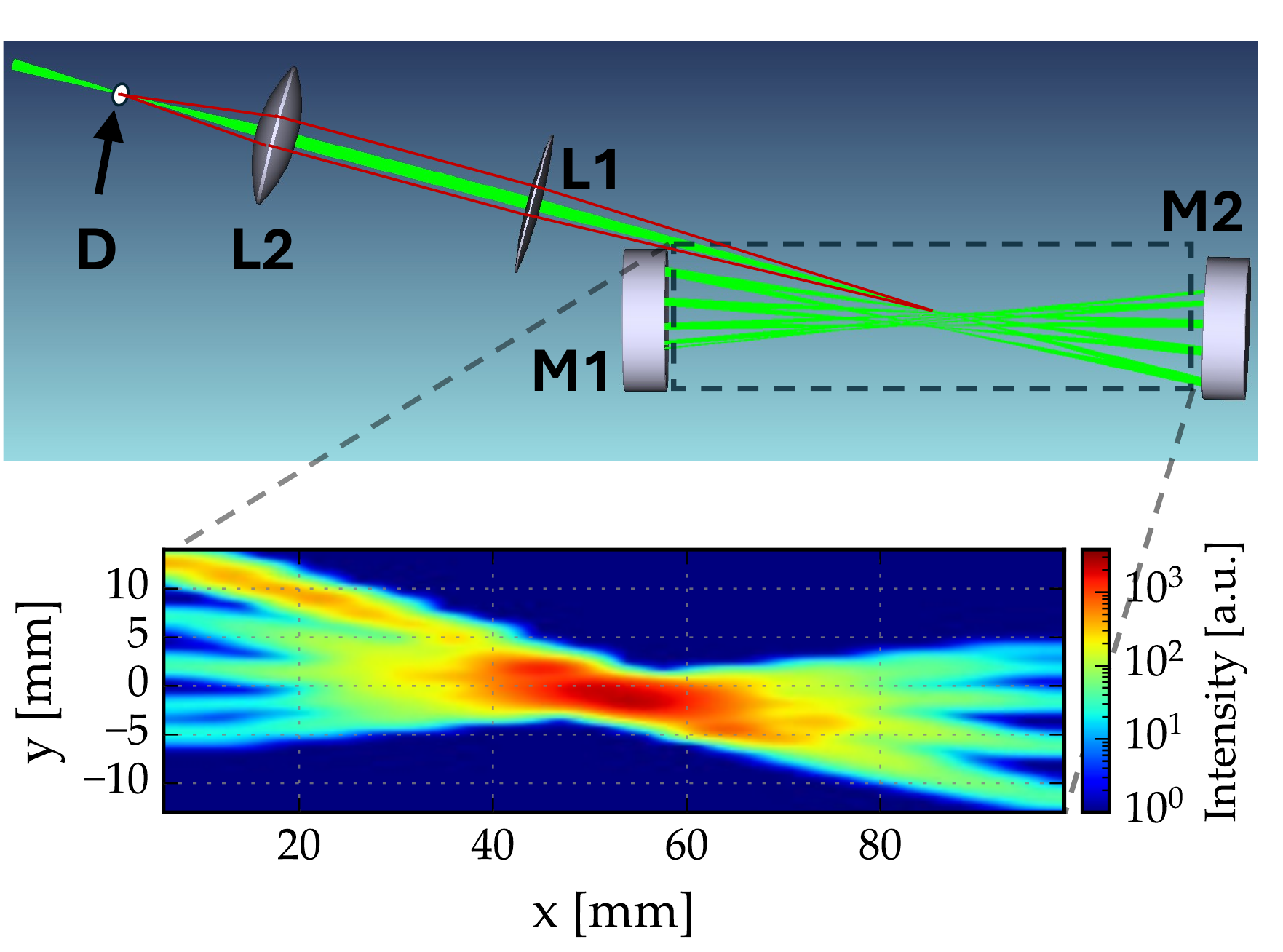}
         \label{subfig: simulation collinear}
     \end{subfigure}
     \hfill
     \begin{subfigure}[b]{0.49\textwidth}
         \centering
         \caption{}
         \includegraphics[width=\textwidth]{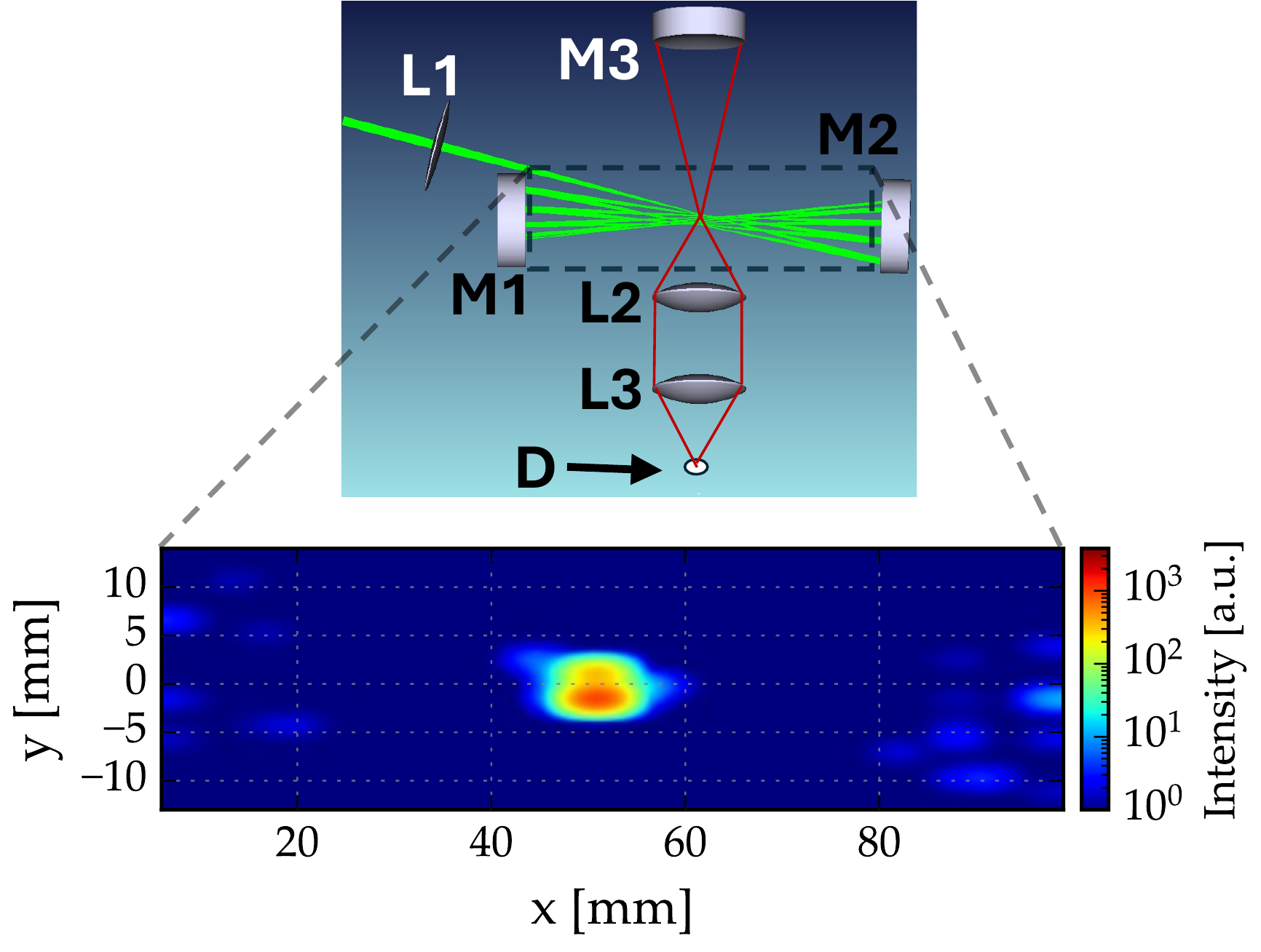}
         \label{subfig: simulation side}
     \end{subfigure}
\setlength{\abovecaptionskip}{0pt} 
 \caption {Evaluation of the collection efficiency in the MPC system with 19 reflections using the optical design and simulation software \textit{Zemax}; comparison of two collection geometries.\\ 
 \subref{subfig: simulation collinear} MPC setup for collinear detection and 2D map of the collected Raman scattered radiation per point within the MPC. \subref{subfig: simulation side} MPC setup for side detection (including feedback mirror M3) and 2D map of the collected Raman scattered radiation per point within the MPC. Red lines indicate the cones of detected Raman signal. Abbreviations: L$_i$: lenses; M$_i$: mirrors; D: detector.}
 \label{fig: simulations}
\end{figure}

A straightforward strategy to reduce fluorescence is to collect Raman light from the side of the cavity to avoid collecting fluorescence generated by optical components along the collinear beam path~\cite{niklas_short_2021, wang_review_2020, yang_highly_2016, yang_high-sensitivity_2024, guo_high-sensitivity_2021, li_near-confocal_2008, miao_parabolic_2024}. However, this approach comes at the cost of lower signal. To assess the penalty in signal collection, ray-tracing based scattering simulations were performed for the two-mirror MPC. \autoref{subfig: simulation collinear} shows a simulated beam pattern (19 reflections) and a two-dimensional map of the collection efficiency across the cavity volume for collinear detection. The map was constructed by placing a small scattering volume at different positions in the cavity and recording how much scattered light reaches the detector plane. As expected, only positions on the beam path contribute. However, the collinear configuration maintains a relatively high collection efficiency across a significant fraction of the cavity. The amount of collected light decreases to about 10\% of the maximum value only beyond distances on the order of centimeters from the center along the beam path. This shows that signal is not only collected from the very center (focal points) of the MPC, as is often believed, and it is also not merely the increase of interaction length that determines the signal strength. In other words, the interaction volume should be increased while simultaneously preserving a high collection efficiency to enhance the effective Raman signals. 

For side detection, the situation changes drastically. In the side-collection simulation (\autoref{subfig: simulation side}), a feedback mirror (M3) is added to effectively double the amount of collected signal, as is usually done in experimental setups. Nevertheless, only a small region near the cavity center contributes significantly to the collected signal because the side-view optics capture only light emitted from a limited volume and do not benefit as much from repeated re-imaging along the beam trajectory. The result is a strong reduction in the total detected Raman signal. Integrating over the 2D collection maps yields a 14.4 times lower total collected signal for the side collection compared to collinear detection. \\
Futhermore, it should be noted that an angular dependence of the Raman effect does not contribute to the reduction in signal. As the polarization of the excitation laser was chosen to be perpendicular to the sketch plane in \autoref{fig: simulations}, each molecule scatters light equally in all directions in the sketch plane~\cite{long_raman_2002}. Angular dependence arises only in the planes perpendicular to the sketch plane. \\
The reduced signal of the side collection geometry can to some extent be compensated by longer measurement times as there is no fluorescence saturating the detector, but such long measurement times are impractical for many applications. Consequently, the remainder of this study employs collinear detection while focusing on fluorescence suppression through a careful choice of components.\\
To conclude this section, the simulation approach described herein is a powerful and relatively simple tool for analyzing and assessing the Raman scattering mechanisms in MPCs. It can be used to evaluate the MPC collection efficiency, accounting for important factors, such as the numerical aperture of the collection optics. Optical simulations can be used to compare and optimize different cavity and collection geometries and investigate the pathways of the collected light. Especially for more complex MPC structures, such as 4-mirror-cavities, similar simulations can provide useful insights and optimization capabilities. 

\subsection{Fluorescence Minimization by Exchange of Components}
\begin{figure}
     \centering
     \begin{subfigure}[b]{0.49\textwidth}
         \centering
         \caption{}
         \includegraphics[width=\textwidth]{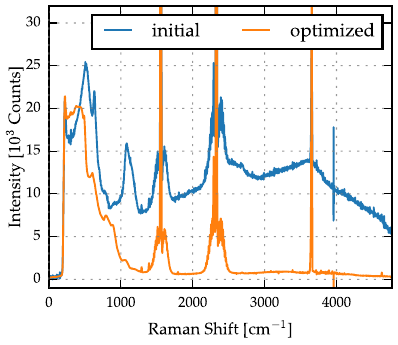}
         \label{subfig: fl final comparison}
     \end{subfigure}
     \hspace{-10pt}
     \begin{subfigure}[b]{0.49\textwidth}
         \centering
         \caption{}
         \includegraphics[width=\textwidth]{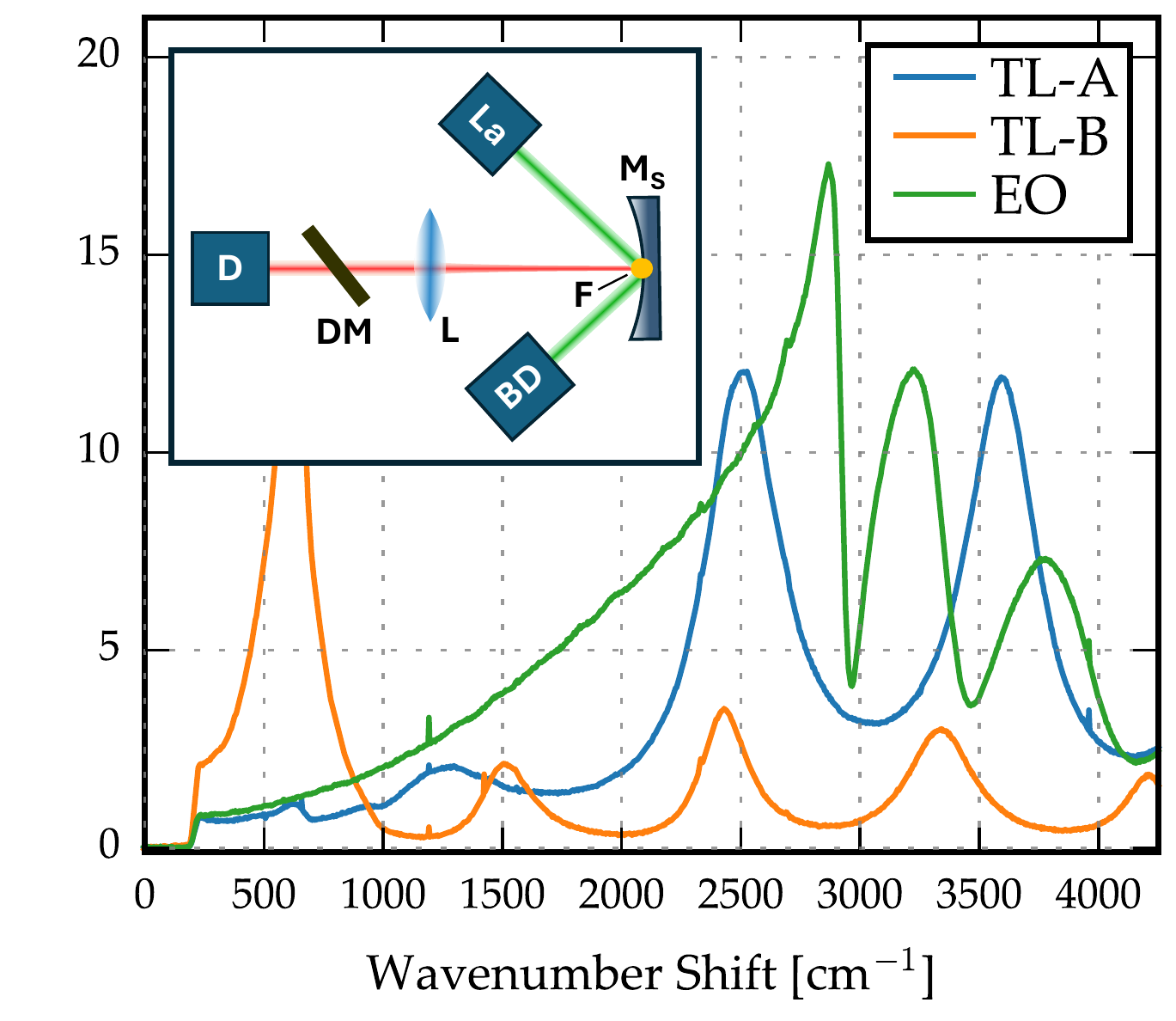}
         \label{subfig: fl mirror comparison}
     \end{subfigure}
\setlength{\abovecaptionskip}{0pt} 
 \caption {Fluorescence background analysis and minimization. \subref{subfig: fl final comparison} Comparison of the fluorescence background of the initial (not optimized) setup and the final fluorescence-minimized setup at an integration time of \SI{10}{s}. \subref{subfig: fl mirror comparison} Comparison of the fluorescence background of different mirrors (TL-A \& TL-B: \textit{Thorlabs} dielectric mirrors from two different batches, EO: dielectric mirror from \textit{Edmund Optics}). The inset shows a sketch of the measurement setup used to obtain these graphs. Abbreviations: La: laser; BD: beam dump; F: focal point of laser and collection optics; M$_{\mathrm{S}}$: sample mirror; L: lens; DM: dichroic mirror; D: detector.}
 \label{fig: fluorescence minimization}
\end{figure}

From \autoref{subfig: spectrum air 500ms}, the fluorescence problem might not be immediately apparent, but the problem becomes clearer when the integration time is increased, e.g. to measure small analyte concentrations. The blue graph in \autoref{subfig: fl final comparison} shows a spectrum of laboratory air with the initial setup (not fluorescence-minimized) and demonstrates that the fluorescence background (and thereby also the noise) already becomes substantial for a still moderate integration time of \SI{10}{s}. \\
With collinear collection, background minimization must target the optical components that the laser is reflected off or intersects, thus generating fluorescence. Considering the MPC setup in \autoref{fig:setup}, all optical components interacting with the beam before it hits the dichroic mirror, can be neglected because the dichroic mirror filters out the red-shifted fluorescence light. This means that only the dichroic mirror itself, the collection lens L1, the window W and the two cavity mirrors M1 and M2 can contribute to the fluorescence background in the signal.
A systematic component survey was performed; the results demonstrated that the anti-reflection-coated N-BK7 lens and window of the initial setup (blue graph in \autoref{subfig: fl final comparison}) contributed to substantial broadband fluorescence, while the dichroic mirror had little influence.\\
Furthermore, it was found out that the dielectric mirrors had a noticeable fluorescence contribution as well. This is depicted in \autoref{subfig: fl mirror comparison}, where the fluorescence backgrounds of \textit{Thorlabs} dielectric mirrors from two different manufacturing batches (TL-A \& TL-B) and of a dielectric mirror from \textit{Edmund Optics} are compared. The \textit{Thorlabs} mirrors have lower background than the \textit{Edmund Optics} mirror (despite the latter having higher reflectance) but there are significant differences between different manufacturing batches of the \textit{Thorlabs} mirrors. According to studies of J. Singh, A. Muller et al., using custom ultra-low-loss mirrors can further reduce the fluorescence background drastically~\cite{muktha_arachchige_portable_2025, singh_high-precision_2023, singh_ambient_2023, singh_isotopic_2021, arachchige_raman_2024}.\\

Hence, the design of the system has been optimized relying on these observations; the following improvements were made to reduce fluorescence: i) the anti-reflection-coated N-BK7 lens and window were replaced by low-fluorescence alternatives consisting of uncoated UV-grade fused silica (UVFS); ii) 6 \textit{Thorlabs} mirrors from various batches were compared and the best-performing were selected; iii) a polarizing beam splitter was introduced in the signal path after the dichroic mirror to suppress the largely unpolarized fluorescence by approximately a factor of two.\\

\autoref{subfig: fl final comparison} shows the resulting fluorescence reduction for representative spectra of laboratory air recorded in otherwise identical conditions (\SI{10}{s} integration time). In the spectral region relevant for many homonuclear diatomic analytes (O$_2$, N$_2$, H$_2$ etc.) (\text{$>$} \SI{1000}{cm^{-1}}), the background level is reduced by roughly a factor of 10 to 15, depending on the specific wavenumber region. This reduction translates directly into an SNR benefit because the dominant noise contribution scales with the baseline signal (see \autoref{subsec: noise analysis}).\\
Therefore, this section demonstrates the importance of a careful selection of materials on the system noise. Components not typically suspected as fluorescence sources, such as high-reflectivity dielectric mirrors, can significantly influence the background and should be tested when building high-sensitivity Raman instruments. Replacing fluorescing optics with non-fluorescing counterparts is a simple, yet often times overlooked method to significantly improve SNR, while preserving the high signal of collinear collection in Raman spectroscopy. With further improvements of this approach, Raman-based gas sensing at the signal shot noise limit is expected to be achievable. 

\subsection{Representative Spectra}
A spectrum of laboratory air for an integration time of \SI{500}{ms} averaged 10 times (\SI{5}{s} total) is shown in \autoref{subfig: spectrum air 500ms} and illustrates the high dynamic range from 80\% (N$_2$) down to \rm{$\sim$}\SI{1000}{ppm} (CO$_2$) of the system (N$_2$ and CO$_2$ have comparable scattering cross sections~\cite{fenner_raman_1973, weber_raman_1979}). The vibrational peaks (Q$_1$ branches) of O$_2$, N$_2$ and H$_2$O are clearly visible and also the Fermi diad of CO$_2$ appears (see \autoref{subsec: noise analysis}).

\begin{figure}
 \centering
 \includegraphics[width=0.99\textwidth]{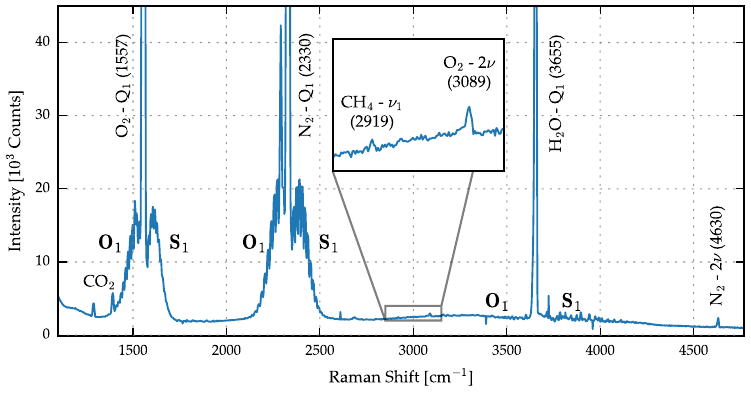}
 \caption{Raman spectrum of laboratory air acquired with \SI{30}{s} integration time averaged six times (\SI{180}{s} total). Longer integration time reveals weaker features (O$_2$ and N$_2$ overtones and peak of CH$_4$) and improves peak quantification.}
 \label{fig: air30s}
\end{figure}

For longer total acquisition, \autoref{fig: air30s} shows a spectrum recorded with \SI{30}{s} integration averaged six times (\SI{180}{s} total). At this increased integration time, the vibrational peaks of O$_2$, N$_2$ and H$_2$O already saturate the detector, but the improved SNR reveals significantly weaker features, such as the rotational O$_1$ and S$_1$ branches of O$_2$, N$_2$ and H$_2$O. \\
Furthermore, the O$_2$ and N$_2$ overtones become observable at \SI{3089}{cm^{-1}} and \SI{4630}{cm^{-1}}, respectively~\cite{zheng_gas_2024, singh_ambient_2023, singh_high-precision_2023, singh_isotopic_2021, velez_spontaneous_2021, petrov_multipass_2016, petrov_multipass_2022, muktha_arachchige_portable_2025, xiao_overtone_2024}.
 A weak band near 2919\,cm$^{-1}$ consistent with ambient CH$_4$~\cite{weber_raman_1979, fenner_raman_1973} can also be detected, which is notable given that typical atmospheric CH$_4$ levels are around \SI{2}{ppm}. These spectra serve as representative examples of the instrument performance measuring an untreated gas mixture (ambient air) and detection limits at a single-digit ppm level.

\subsection{Calibration Curves and Detection Limits}
Calibration measurements were performed for O$_2$ diluted in N$_2$, N$_2$ diluted in O$_2$ and H$_2$ diluted in N$_2$/O$_2$. For each concentration point, six spectra with \SI{30}{s} integration time were recorded and averaged (\SI{180}{s} total). An exemplary spectrum, also showing the Raman peaks of H$_2$, measured with these settings can be found in the \hyperref[sec: SI]{SI} (Figure S4). Raman peaks were quantified by the integral of corresponding Gaussian fits and calibration curves were obtained by weighted linear regression. For O$_2$ and N$_2$ the Q$_1$-branches at \SI{1557}{cm^{-1}} and \SI{2330}{cm^{-1}}, respectively, were used for quantification and for H$_2$, the Q$_1$(1) at \SI{4157}{cm^{-1}}~\cite{weber_raman_1979, fenner_raman_1973, wang_cavity-enhanced_2023} was used because of its prominence and the low background signal in this spectral region. Exemplary graphs showing the quality of the Gaussian fits for O$_2$, N$_2$, and H$_2$ can be found in the \hyperref[sec: SI]{SI} (Figure S5).\\

\begin{figure}
     \centering
     \begin{subfigure}[b]{0.33\textwidth}
         \centering
         \caption{Calibration of O$_2$}
         \includegraphics[width=\textwidth]{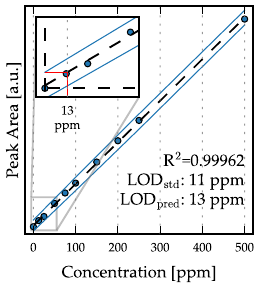}
         \label{subfig: calibration O2}
     \end{subfigure}
     \hspace{-7pt}
     \begin{subfigure}[b]{0.33\textwidth}
         \centering
         \caption{Calibration of N$_2$}
         \includegraphics[width=\textwidth]{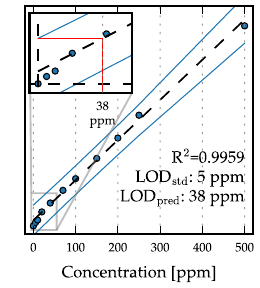}
         \label{subfig: calibration N2}
     \end{subfigure}
     \hspace{-7pt}
     \begin{subfigure}[b]{0.33\textwidth}
         \centering
         \caption{Calibration of H$_2$}
         \includegraphics[width=\textwidth]{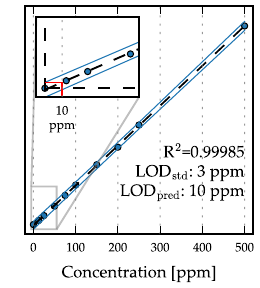}
         \label{subfig: calibration H2}
     \end{subfigure}
\setlength{\abovecaptionskip}{0pt} 
 \caption {Calibration Curves for \subref{subfig: calibration O2} O$_2$ using the Q$_1$ branch at \SI{1557}{cm^{-1}} \subref{subfig: calibration N2} N$_2$ using the Q$_1$ branch at \SI{2330}{cm^{-1}} and \subref{subfig: calibration H2} H$_2$ using the Q$_1$(1) transition at \SI{4157}{cm^{-1}} for quantification (see also Figures S4 and S5 in the \hyperref[sec: SI]{SI}). The insets show how LOD$_{\mathrm{pred}}$ is obtained from the prediction bands (blue) of the calibration. The signal (peak area, represented by blue circles) is obtained by a Gaussian fit. Each measurement point is an average of six measurements with an integration time of \SI{30}{s} (\SI{180}{s} total).}
 \label{fig: calibratino curves}
\end{figure}

Figures \ref{subfig: calibration O2}--\ref{subfig: calibration H2} show the resulting calibration curves including 99.7\% prediction bands. The O$_2$ and H$_2$ calibrations show excellent linearity with coefficients of determination $R^2>0.9996$ and detection limits of \SI{11}{ppm} to \SI{13}{ppm} for O$_2$ and \SI{3}{ppm} to \SI{10}{ppm} for H$_2$. The low LOD for H$_2$ can be explained by the non-existing H$_2$ contamination from outside the gas cell. 
For N$_2$ in O$_2$ a slight curvature is observed, which is attributed to a drifting N$_2$ background caused by imperfect sealing of the 3D-printed cell material. This is also reflected in the detection limit range of \SI{5}{ppm} to \SI{38}{ppm} for N$_2$, which highlights the importance of considering different definitions for the detection limit: the blank sample in the N$_2$ calibration has a very low standard deviation. This yields a very low detection limit of \SI{5}{ppm} according to the standard IUPAC-style definition given in \autoref{eq: iupac}. The more sophisticated definition using the prediction bands (\autoref{eq: LOD_pred}) considers the uncertainty of the whole calibration procedure and yields a much more realistic detection limit of \SI{38}{ppm} for this calibration.
Table \ref{tab:lod} summarizes the limits of detection calculated using both LOD definitions described above. For a \SI{180}{s} measurement time, LOD values are in the few-ppm to few-tens-of-ppm range depending on gas and LOD definition.\\
A way to improve these LOD values would be to manufacture the gas cell out of metal and introduce gas-tight O-ring sealing elements. This would drastically reduce the air contamination and significantly improve the LOD of N$_2$. Furthermore, this cell could be designed to withstand sample gas pressures of a few bar to further decrease the LOD values. Previous studies have shown that SNR and LOD scale roughly linearly with the sample gas pressure~\cite{wang_cavity-enhanced_2023, singh_ambient_2023, singh_high-precision_2023, velez_spontaneous_2021, arachchige_raman_2024, wang_highly_2023, petrov_multipass_2022, muktha_arachchige_portable_2025}, because $\rho$ in \autoref{eq: intensity} increases linearly with pressure while the noisy fluorescence background, stemming from the optics, remains constant.

\begin{table}
\caption{Detection limits (ppm) and calibration quality metrics for a 180\,s total measurement time. $\mathrm{LOD}_{\mathrm{std}}$: standard 3$\sigma$ definition; $\mathrm{LOD}_{\mathrm{pred}}$: prediction-band based.}
\centering
\begin{tabular}{lcccc}
\toprule
Gas & $\mathrm{LOD}_{\mathrm{std}}$ [ppm] & $\mathrm{LOD}_{\mathrm{pred}}$ [ppm] & $R^2$ & Diluent Gases \\
\midrule
O$_2$ & 11 & 13 & 0.99962 &  N$_2$\\
N$_2$ & 5 & 38 & 0.99590 & O$_2$\\
H$_2$ & 3 & 10 & 0.99985 & N$_2$, O$_2$ \\
\bottomrule
\end{tabular}
\label{tab:lod}
\end{table}

\newpage
\section{Conclusion and Outlook}
A compact gas sensor based on cavity-enhanced Raman spectroscopy (CERS) using a non-resonant two-mirror multi-pass cavity (MPC) has been demonstrated. The central objective of this work was to establish fluorescence minimization as a practical route towards high-sensitivity Raman gas spectroscopy in a robust, field-oriented instrumental concept. To this end, we combined (i) a quantitative noise analysis, (ii) numerical scattering simulations to assess collection geometries, and (iii) targeted optical-design and filtering measures to suppress background while preserving the MPC signal gain. 

A CCD-specific noise model explicitly linking the fluorescence baseline to measurement noise was employed to motivate fluorescence suppression at the design stage, rather than relying primarily on baseline removal in post-processing. Ray-tracing-based Raman scattering simulations further showed that, for a two-mirror MPC, side collection substantially restricts the effective collection volume and leads to a pronounced loss in detected signal (14.4$\times$ lower compared to collinear collection in the simulated geometry). Consequently, side collection is not a suitable fluorescence-mitigation strategy when high MPC enhancement and short measurement times are required. Fluorescence contributions from individual optical components were therefore addressed directly. By replacing fluorescent components (including cavity mirrors) and introducing polarization filtering, the fluorescence background in the relevant spectral region was reduced by roughly one order of magnitude, translating into a marked improvement in SNR and LOD. The capabilities of the fluorescence-minimized system were demonstrated in representative measurements of laboratory air, where even weak Raman features were resolved, including ambient CH$_4$ (\rm{$\sim$}\SI{2}{ppm}). Calibration curves for O$_2$, N$_2$ and H$_2$ yielded detection limits (depending on the exact definition) of 11--\SI{13}{ppm}, 5--\SI{38}{ppm} and 3--\SI{10}{ppm}, respectively, at a total measurement time of \SI{180}{s}. 

Overall, these results identify fluorescence management as a key design principle for robust, high-sensitivity MPC-based CERS instrumentation. Several clear upgrade paths emerge from the present analysis: (i) further reduction of background by selecting even less-fluorescent optical components, especially mirrors; (ii) improving gas integrity by replacing the 3D-printed cell with a metallic gas cell employing gas-tight O-ring sealing to prevent N$_2$ and O$_2$ contamination; and (iii) leveraging moderate sample-gas pressurization to amplify Raman signals and thereby reduce LODs. On the detection side, higher optical throughput can be achieved by using a spectrometer with a larger slit, and/or by increasing excitation power (\SI{532}{nm} CW lasers with optical powers of several Watts are readily available). With these advancements, sub-ppm-level detection limits with measurement times in the seconds range are expected.

\section*{Supplementary information}

Link to Supplementary Information.
\label{sec: SI}

\section*{Funding}

Financial support by the Austrian Research Promotion Agency FFG and the state of Upper
Austria via the project H2lytics (project number: FO999909473) as part of the call Future
Energy Technologies Ausschreibung 2023 is acknowledged.

\section*{Roles}

S.H-R.: Conceptualization, Data curation, Formal analysis, Investigation, Methodology, Software, Validation, Visualization, Writing - original draft.\newline
R.Z.: Conceptualization, Formal analysis, Funding acquisition, Investigation, Methodology, Project administration, Supervision, Writing - review \& editing.\newline
P.G.: Conceptualization, Funding acquisition, Methodology, Writing - review \& editing.\newline
I.Z.: Conceptualization, Methodology, Writing - review \& editing.\newline
J.D.P.: Formal analysis, Methodology, Supervision, Writing - review \& editing. \newline
M.B.: Conceptualization, Formal analysis, Funding acquisition, Methodology, Project administration, Resources, Supervision, Writing - review \& editing.\newline
All authors approved the final manuscript.

\section*{Data}

The data that support the findings of this study are available from the corresponding author upon reasonable request.

\section*{Conflict of Interest}
There are no conflicts of interest.


\section*{References}
\printbibliography[heading=none]

@article{niklas_short_2021,
	title = {A Short Review of Cavity-Enhanced Raman Spectroscopy for Gas Analysis},
	volume = {21},
	rights = {https://creativecommons.org/licenses/by/4.0/},
	issn = {1424-8220},
	url = {https://www.mdpi.com/1424-8220/21/5/1698},
	doi = {10.3390/s21051698},
	pages = {1698},
	number = {5},
	journaltitle = {Sensors},
	shortjournal = {Sensors},
	author = {Niklas, Christian and Wackerbarth, Hainer and Ctistis, Georgios},
	urldate = {2024-06-25},
	date = {2021-03-02},
	langid = {english},
}

@article{yang_multiple_2023,
	title = {Multiple Gas Detection by Cavity-Enhanced Raman Spectroscopy with Sub-ppm Sensitivity},
	volume = {95},
	rights = {https://doi.org/10.15223/policy-029},
	issn = {0003-2700, 1520-6882},
	url = {https://pubs.acs.org/doi/10.1021/acs.analchem.2c05432},
	doi = {10.1021/acs.analchem.2c05432},
	pages = {5652--5660},
	number = {13},
	journaltitle = {Analytical Chemistry},
	shortjournal = {Anal. Chem.},
	author = {Yang, Qing-ying and Tan, Yan and Qu, Zi-han and Sun, Yu and Liu, An-wen and Hu, Shui-ming},
	urldate = {2024-06-25},
	date = {2023-04-04},
	langid = {english},
}

@article{wang_multigas_2020,
	title = {Multigas Analysis by Cavity-Enhanced Raman Spectroscopy for Power Transformer Diagnosis},
	volume = {92},
	rights = {https://doi.org/10.15223/policy-029},
	issn = {0003-2700, 1520-6882},
	url = {https://pubs.acs.org/doi/10.1021/acs.analchem.0c00179},
	doi = {10.1021/acs.analchem.0c00179},
	pages = {5969--5977},
	number = {8},
	journaltitle = {Analytical Chemistry},
	shortjournal = {Anal. Chem.},
	author = {Wang, Pinyi and Chen, Weigen and Wang, Jianxin and Tang, Jun and Shi, Yongli and Wan, Fu},
	urldate = {2024-06-25},
	date = {2020-04-21},
	langid = {english},
}

@article{wang_cavity-enhanced_2023,
	title = {Cavity-Enhanced Raman Spectroscopy for Detection of Trace Gaseous Impurities in Hydrogen for Fuel Cells},
	volume = {95},
	rights = {https://doi.org/10.15223/policy-029},
	issn = {0003-2700, 1520-6882},
	url = {https://pubs.acs.org/doi/10.1021/acs.analchem.3c00066},
	doi = {10.1021/acs.analchem.3c00066},
	pages = {6894--6904},
	number = {17},
	journaltitle = {Analytical Chemistry},
	shortjournal = {Anal. Chem.},
	author = {Wang, Pinyi and Chen, Weigen and Wang, Jianxin and Lu, Yongkang and Tang, Zijie and Tan, Yaxiong},
	urldate = {2024-10-21},
	date = {2023-05-02},
	langid = {english},
}

@book{long_raman_2002,
	edition = {1},
	title = {The Raman Effect: A Unified Treatment of the Theory of Raman Scattering by Molecules},
	rights = {http://doi.wiley.com/10.1002/tdm\_license\_1.1},
	isbn = {0-470-84576-7},
	url = {https://onlinelibrary.wiley.com/doi/book/10.1002/0470845767},
	doi = {10.1002/0470845767},
	shorttitle = {The Raman Effect},
	publisher = {Wiley},
	author = {Long, Derek A.},
	urldate = {2024-10-22},
	date = {2002-04-15},
	langid = {english},
}

@article{fenner_raman_1973,
	title = {Raman cross section of some simple gases},
	volume = {63},
	rights = {https://doi.org/10.1364/{OA}\_License\_v1\#{VOR}},
	issn = {0030-3941},
	url = {https://opg.optica.org/abstract.cfm?URI=josa-63-1-73},
	doi = {10.1364/JOSA.63.000073},
	pages = {73},
	number = {1},
	journaltitle = {Journal of the Optical Society of America},
	shortjournal = {J. Opt. Soc. Am.},
	author = {Fenner, Wayne R. and Hyatt, Howard A. and Kellam, John M. and Porto, S. P. S.},
	urldate = {2024-11-12},
	date = {1973-01-01},
	langid = {english},
}

@collection{weber_raman_1979,
	location = {Berlin, Heidelberg},
	title = {Raman Spectroscopy of Gases and Liquids},
	volume = {11},
	rights = {http://www.springer.com/tdm},
	isbn = {978-3-642-81279-8},
	url = {http://link.springer.com/10.1007/978-3-642-81279-8},
	doi = {10.1007/978-3-642-81279-8},
	series = {Topics in Current Physics},
	publisher = {Springer Berlin Heidelberg},
	editor = {Weber, Alfons},
	urldate = {2024-11-12},
	date = {1979},
	langid = {english},
}

@article{zheng_gas_2024,
	title = {Gas detection using cavity-enhanced Raman spectroscopy with a bidirectional multi-pass cell and polarization beam-splitting optical path},
	volume = {130},
	issn = {0946-2171, 1432-0649},
	url = {https://link.springer.com/10.1007/s00340-024-08285-y},
	doi = {10.1007/s00340-024-08285-y},
	pages = {144},
	number = {8},
	journaltitle = {Applied Physics B},
	shortjournal = {Appl. Phys. B},
	author = {Zheng, Yuhao and Zou, Xiaer and He, Sailing},
	urldate = {2024-11-18},
	date = {2024-08},
	langid = {english},
}

@article{yang_high-sensitivity_2024,
	title = {High-Sensitivity and In Situ Multi-Component Detection of Gases Based on Multiple-Reflection-Cavity-Enhanced Raman Spectroscopy},
	volume = {24},
	rights = {https://creativecommons.org/licenses/by/4.0/},
	issn = {1424-8220},
	url = {https://www.mdpi.com/1424-8220/24/17/5825},
	doi = {10.3390/s24175825},
	pages = {5825},
	number = {17},
	journaltitle = {Sensors},
	shortjournal = {Sensors},
	author = {Yang, Dewang and Li, Wenhua and Tian, Haoyue and Chen, Zhigao and Ji, Yuhang and Dong, Hui and Wang, Yongmei},
	urldate = {2024-11-18},
	date = {2024-09-07},
	langid = {english},
}

@article{petrov_multipass_2016,
	title = {Multipass optical system for a Raman gas spectrometer},
	volume = {55},
	rights = {https://doi.org/10.1364/{OA}\_License\_v1\#{VOR}},
	issn = {0003-6935, 1539-4522},
	url = {https://opg.optica.org/abstract.cfm?URI=ao-55-33-9521},
	doi = {10.1364/AO.55.009521},
	pages = {9521},
	number = {33},
	journaltitle = {Applied Optics},
	shortjournal = {Appl. Opt.},
	author = {Petrov, Dmitry V.},
	urldate = {2024-12-12},
	date = {2016-11-20},
	langid = {english},
}

@article{petrov_multipass_2022,
	title = {Multipass Raman gas analyzer for monitoring of atmospheric air composition},
	volume = {152},
	issn = {00303992},
	url = {https://linkinghub.elsevier.com/retrieve/pii/S0030399222003127},
	doi = {10.1016/j.optlastec.2022.108155},
	pages = {108155},
	journaltitle = {Optics \& Laser Technology},
	shortjournal = {Optics \& Laser Technology},
	author = {Petrov, D.V. and Matrosov, I.I. and Kostenko, M.A.},
	urldate = {2024-12-12},
	date = {2022-08},
	langid = {english},
}

@article{gomez_velez_trace_2020,
	title = {Trace gas sensing using diode-pumped collinearly detected spontaneous Raman scattering enhanced by a multipass cell},
	volume = {45},
	issn = {0146-9592, 1539-4794},
	url = {https://opg.optica.org/abstract.cfm?URI=ol-45-1-133},
	doi = {10.1364/OL.45.000133},
	pages = {133},
	number = {1},
	journaltitle = {Optics Letters},
	shortjournal = {Opt. Lett.},
	author = {Gomez Velez, Juan and Muller, Andreas},
	urldate = {2024-12-12},
	date = {2020-01-01},
	langid = {english},
}

@article{guo_high-sensitivity_2021,
	title = {High-Sensitivity Raman Gas Probe for In Situ Multi-Component Gas Detection},
	volume = {21},
	rights = {https://creativecommons.org/licenses/by/4.0/},
	issn = {1424-8220},
	url = {https://www.mdpi.com/1424-8220/21/10/3539},
	doi = {10.3390/s21103539},
	pages = {3539},
	number = {10},
	journaltitle = {Sensors},
	shortjournal = {Sensors},
	author = {Guo, Jinjia and Luo, Zhao and Liu, Qingsheng and Yang, Dewang and Dong, Hui and Huang, Shuke and Kong, Andong and Wu, Lulu},
	urldate = {2024-12-12},
	date = {2021-05-19},
	langid = {english},
}

@article{wang_review_2020,
	title = {A review of cavity-enhanced Raman spectroscopy as a gas sensing method},
	volume = {55},
	issn = {0570-4928, 1520-569X},
	url = {https://www.tandfonline.com/doi/full/10.1080/05704928.2019.1661850},
	doi = {10.1080/05704928.2019.1661850},
	pages = {393--417},
	number = {5},
	journaltitle = {Applied Spectroscopy Reviews},
	shortjournal = {Applied Spectroscopy Reviews},
	author = {Wang, Pinyi and Chen, Weigen and Wan, Fu and Wang, Jianxin and Hu, Jin},
	urldate = {2024-12-12},
	date = {2020-05-27},
	langid = {english},
}

@article{xiao-yun_diagnosis_2008,
	title = {Diagnosis of Multiple Gases Separated from Transformer Oil Using Cavity-Enhanced Raman Spectroscopy},
	volume = {25},
	issn = {0256-307X, 1741-3540},
	url = {https://iopscience.iop.org/article/10.1088/0256-307X/25/9/062},
	doi = {10.1088/0256-307X/25/9/062},
	pages = {3326--3329},
	number = {9},
	journaltitle = {Chinese Physics Letters},
	shortjournal = {Chinese Phys. Lett.},
	author = {Xiao-Yun, Li and Yu-Xing, Xia and Ju-Ming, Huang and Li, Zhan},
	urldate = {2024-12-31},
	date = {2008-09},
	langid = {english},
}

@article{gebicki_application_2016,
	title = {Application of electrochemical sensors and sensor matrixes for measurement of odorous chemical compounds},
	volume = {77},
	issn = {01659936},
	url = {https://linkinghub.elsevier.com/retrieve/pii/S0165993615300923},
	doi = {10.1016/j.trac.2015.10.005},
	pages = {1--13},
	journaltitle = {{TrAC} Trends in Analytical Chemistry},
	shortjournal = {{TrAC} Trends in Analytical Chemistry},
	author = {Gebicki, Jacek},
	urldate = {2025-01-02},
	date = {2016-03},
	langid = {english},
}

@article{wilson_applications_2009,
	title = {Applications and Advances in Electronic-Nose Technologies},
	volume = {9},
	rights = {https://creativecommons.org/licenses/by/3.0/},
	issn = {1424-8220},
	url = {https://www.mdpi.com/1424-8220/9/7/5099},
	doi = {10.3390/s90705099},
	pages = {5099--5148},
	number = {7},
	journaltitle = {Sensors},
	shortjournal = {Sensors},
	author = {Wilson, Alphus  D. and Baietto, Manuela},
	urldate = {2025-01-02},
	date = {2009-06-29},
	langid = {english},
}

@article{bartle_history_2002,
	title = {History of gas chromatography},
	volume = {21},
	rights = {https://www.elsevier.com/tdm/userlicense/1.0/},
	issn = {01659936},
	url = {https://linkinghub.elsevier.com/retrieve/pii/S0165993602008063},
	doi = {10.1016/S0165-9936(02)00806-3},
	pages = {547--557},
	number = {9},
	journaltitle = {{TrAC} Trends in Analytical Chemistry},
	shortjournal = {{TrAC} Trends in Analytical Chemistry},
	author = {Bartle, Keith D. and Myers, Peter},
	urldate = {2025-01-02},
	date = {2002-09},
	langid = {english},
}

@article{kaminski_determination_2003,
	title = {Determination of carbon monoxide, methane and carbon dioxide in refinery hydrogen gases and air by gas chromatography},
	volume = {989},
	rights = {https://www.elsevier.com/tdm/userlicense/1.0/},
	issn = {00219673},
	url = {https://linkinghub.elsevier.com/retrieve/pii/S0021967303000323},
	doi = {10.1016/S0021-9673(03)00032-3},
	pages = {277--283},
	number = {2},
	journaltitle = {Journal of Chromatography A},
	shortjournal = {Journal of Chromatography A},
	author = {Kamiński, Marian and Kartanowicz, Rafał and Jastrzębski, Daniel and Kamiński, Marcin M.},
	urldate = {2025-01-02},
	date = {2003-03},
	langid = {english},
}

@article{van_ruth_methods_2001,
	title = {Methods for gas chromatography-olfactometry: a review},
	volume = {17},
	rights = {https://www.elsevier.com/tdm/userlicense/1.0/},
	issn = {13890344},
	url = {https://linkinghub.elsevier.com/retrieve/pii/S1389034401000703},
	doi = {10.1016/S1389-0344(01)00070-3},
	shorttitle = {Methods for gas chromatography-olfactometry},
	pages = {121--128},
	number = {4},
	journaltitle = {Biomolecular Engineering},
	shortjournal = {Biomolecular Engineering},
	author = {Van Ruth, Saskia M},
	urldate = {2025-01-02},
	date = {2001-05},
	langid = {english},
}

@article{fu_recent_2022,
	title = {Recent progress on laser absorption spectroscopy for determination of gaseous chemical species},
	volume = {57},
	issn = {0570-4928, 1520-569X},
	url = {https://www.tandfonline.com/doi/full/10.1080/05704928.2020.1857258},
	doi = {10.1080/05704928.2020.1857258},
	pages = {112--152},
	number = {2},
	journaltitle = {Applied Spectroscopy Reviews},
	shortjournal = {Applied Spectroscopy Reviews},
	author = {Fu, Bo and Zhang, Chenghong and Lyu, Wenhao and Sun, Jingxuan and Shang, Ce and Cheng, Yuan and Xu, Lijun},
	urldate = {2025-01-02},
	date = {2022-02-07},
	langid = {english},
}

@article{okeefe_cavity_1988,
	title = {Cavity ring-down optical spectrometer for absorption measurements using pulsed laser sources},
	volume = {59},
	issn = {0034-6748, 1089-7623},
	url = {https://pubs.aip.org/rsi/article/59/12/2544/313248/Cavity-ring-down-optical-spectrometer-for},
	doi = {10.1063/1.1139895},
	pages = {2544--2551},
	number = {12},
	journaltitle = {Review of Scientific Instruments},
	author = {O’Keefe, Anthony and Deacon, David A. G.},
	urldate = {2025-01-02},
	date = {1988-12-01},
	langid = {english},
}

@article{kreuzer_ultralow_1971,
	title = {Ultralow Gas Concentration Infrared Absorption Spectroscopy},
	volume = {42},
	issn = {0021-8979},
	url = {https://doi.org/10.1063/1.1660651},
	doi = {10.1063/1.1660651},
	pages = {2934--2943},
	number = {7},
	journaltitle = {Journal of Applied Physics},
	shortjournal = {Journal of Applied Physics},
	author = {Kreuzer, L. B.},
	urldate = {2025-01-02},
	date = {1971-06-01},
}

@article{souza_filho_green_2022,
	title = {Green steel at its crossroads: Hybrid hydrogen-based reduction of iron ores},
	volume = {340},
	issn = {09596526},
	url = {https://linkinghub.elsevier.com/retrieve/pii/S0959652622004437},
	doi = {10.1016/j.jclepro.2022.130805},
	shorttitle = {Green steel at its crossroads},
	pages = {130805},
	journaltitle = {Journal of Cleaner Production},
	shortjournal = {Journal of Cleaner Production},
	author = {Souza Filho, Isnaldi R. and Springer, Hauke and Ma, Yan and Mahajan, Ankita and Da Silva, Cauê C. and Kulse, Michael and Raabe, Dierk},
	urldate = {2025-01-02},
	date = {2022-03},
	langid = {english},
}

@article{miao_parabolic_2024,
	title = {Parabolic mirror cavity-enhanced Raman spectroscopy for trace gas detection},
	volume = {49},
	issn = {0146-9592, 1539-4794},
	url = {https://opg.optica.org/abstract.cfm?URI=ol-49-19-5455},
	doi = {10.1364/OL.534842},
	pages = {5455},
	number = {19},
	journaltitle = {Optics Letters},
	shortjournal = {Opt. Lett.},
	author = {Miao, Junfang and Liu, Jiaxiang and Ning, Zhiqiang and Xu, Haichun and Pan, Ying and Li, Zhengang and Fang, Yonghua},
	urldate = {2025-03-03},
	date = {2024-10-01},
	langid = {english},
}

@article{yang_highly_2016,
	title = {Highly sensitive Raman system for dissolved gas analysis in water},
	volume = {55},
	rights = {https://doi.org/10.1364/{OA}\_License\_v1\#{VOR}},
	issn = {0003-6935, 1539-4522},
	url = {https://opg.optica.org/abstract.cfm?URI=ao-55-27-7744},
	doi = {10.1364/AO.55.007744},
	pages = {7744},
	number = {27},
	journaltitle = {Applied Optics},
	shortjournal = {Appl. Opt.},
	author = {Yang, Dewang and Guo, Jinjia and Liu, Qingsheng and Luo, Zhao and Yan, Jingwen and Zheng, Ronger},
	urldate = {2025-03-03},
	date = {2016-09-20},
	langid = {english},
}

@article{li_near-confocal_2008,
	title = {Near-confocal cavity-enhanced Raman spectroscopy for multitrace-gas detection},
	volume = {33},
	rights = {https://doi.org/10.1364/{OA}\_License\_v1\#{VOR}},
	issn = {0146-9592, 1539-4794},
	url = {https://opg.optica.org/abstract.cfm?URI=ol-33-18-2143},
	doi = {10.1364/OL.33.002143},
	pages = {2143},
	number = {18},
	journaltitle = {Optics Letters},
	shortjournal = {Opt. Lett.},
	author = {Li, Xiaoyun and Xia, Yuxing and Zhan, Li and Huang, Juming},
	urldate = {2025-03-03},
	date = {2008-09-15},
	langid = {english},
}

@article{black_introduction_2001,
	title = {An introduction to Pound–Drever–Hall laser frequency stabilization},
	volume = {69},
	issn = {0002-9505, 1943-2909},
	url = {https://pubs.aip.org/ajp/article/69/1/79/1055569/An-introduction-to-Pound-Drever-Hall-laser},
	doi = {10.1119/1.1286663},
	pages = {79--87},
	number = {1},
	journaltitle = {American Journal of Physics},
	author = {Black, Eric D.},
	urldate = {2025-05-05},
	date = {2001-01-01},
	langid = {english},
}

@article{nickerson_review_2019,
	title = {A review of Pound-Drever-Hall laser frequency locking},
	url = {https://jila1.nickersonm.com/papers/PDH%20Locking%20Review.pdf},
	pages = {7},
	journaltitle = {{JILA}, University of Colorado and {NIST}},
	author = {Nickerson, M},
	urldate = {2025-11-17},
	date = {2019},
	langid = {english},
}

@article{wang_highly_2023,
	title = {Highly sensitive multi-pass cavity enhanced Raman spectroscopy with novel polarization filtering for quantitative measurement of {SF}6 decomposed components in gas-insulated power equipment},
	volume = {380},
	issn = {09254005},
	url = {https://linkinghub.elsevier.com/retrieve/pii/S0925400523000655},
	doi = {10.1016/j.snb.2023.133350},
	pages = {133350},
	journaltitle = {Sensors and Actuators B: Chemical},
	shortjournal = {Sensors and Actuators B: Chemical},
	author = {Wang, Jianxin and Wang, Pinyi and Chen, Weigen and Wan, Fu and Lu, Yongkang and Tang, Zijie and Dong, Anning and Lei, Zemin and Zhang, Zhixian},
	urldate = {2025-06-10},
	date = {2023-04},
	langid = {english},
}

@article{muktha_arachchige_portable_2025,
	title = {Portable Raman hydrogen concentration mapping with parts-per-billion sensitivity},
	volume = {64},
	rights = {https://doi.org/10.1364/{OA}\_License\_v2\#{AM}},
	issn = {1559-128X, 2155-3165},
	url = {https://opg.optica.org/abstract.cfm?URI=ao-64-13-3646},
	doi = {10.1364/ao.558965},
	pages = {3646},
	number = {13},
	journaltitle = {Applied Optics},
	shortjournal = {Appl. Opt.},
	publisher = {Optica Publishing Group},
	author = {Muktha Arachchige, Charuka and Muller, Andreas},
	urldate = {2025-07-21},
	date = {2025-05-01},
	langid = {english},
}

@article{mermet_calibration_2010,
	title = {Calibration in atomic spectrometry: A tutorial review dealing with quality criteria, weighting procedures and possible curvatures},
	volume = {65},
	rights = {https://www.elsevier.com/tdm/userlicense/1.0/},
	issn = {05848547},
	url = {https://linkinghub.elsevier.com/retrieve/pii/S0584854710001394},
	doi = {10.1016/j.sab.2010.05.007},
	shorttitle = {Calibration in atomic spectrometry},
	pages = {509--523},
	number = {7},
	journaltitle = {Spectrochimica Acta Part B: Atomic Spectroscopy},
	shortjournal = {Spectrochimica Acta Part B: Atomic Spectroscopy},
	author = {Mermet, Jean-Michel},
	urldate = {2025-08-04},
	date = {2010-07},
	langid = {english},
}

@article{mermet_limit_2008,
	title = {Limit of quantitation in atomic spectrometry: An unambiguous concept?},
	volume = {63},
	rights = {https://www.elsevier.com/tdm/userlicense/1.0/},
	issn = {05848547},
	url = {https://linkinghub.elsevier.com/retrieve/pii/S0584854707004430},
	doi = {10.1016/j.sab.2007.11.029},
	shorttitle = {Limit of quantitation in atomic spectrometry},
	pages = {166--182},
	number = {2},
	journaltitle = {Spectrochimica Acta Part B: Atomic Spectroscopy},
	shortjournal = {Spectrochimica Acta Part B: Atomic Spectroscopy},
	author = {Mermet, Jean-Michel},
	urldate = {2025-08-04},
	date = {2008-02},
	langid = {english},
}

@article{irie_technique_2008,
	title = {A Technique for Evaluation of {CCD} Video-Camera Noise},
	volume = {18},
	rights = {https://ieeexplore.ieee.org/Xplorehelp/downloads/license-information/{IEEE}.html},
	issn = {1051-8215, 1558-2205},
	url = {http://ieeexplore.ieee.org/document/4400030/},
	doi = {10.1109/TCSVT.2007.913972},
	pages = {280--284},
	number = {2},
	journaltitle = {{IEEE} Transactions on Circuits and Systems for Video Technology},
	shortjournal = {{IEEE} Trans. Circuits Syst. Video Technol.},
	author = {Irie, K. and {McKinnon}, A.E. and Unsworth, K. and Woodhead, I.M.},
	urldate = {2025-08-08},
	date = {2008-02},
	langid = {english},
}

@article{singh_isotopic_2021,
	title = {Isotopic trace analysis of water vapor with multipass cavity Raman scattering},
	volume = {146},
	issn = {0003-2654, 1364-5528},
	url = {https://xlink.rsc.org/?DOI=D1AN01254A},
	doi = {10.1039/D1AN01254A},
	pages = {6482--6489},
	number = {21},
	journaltitle = {The Analyst},
	shortjournal = {Analyst},
	author = {Singh, Jaspreet and Muller, Andreas},
	urldate = {2025-09-08},
	date = {2021},
	langid = {english},
}

@article{long_limit_1983,
	title = {Limit of Detection A Closer Look at the {IUPAC} Definition},
	volume = {55},
	issn = {0003-2700, 1520-6882},
	url = {https://pubs.acs.org/doi/abs/10.1021/ac00258a724},
	doi = {10.1021/ac00258a724},
	pages = {712A--724A},
	number = {7},
	journaltitle = {Analytical Chemistry},
	shortjournal = {Anal. Chem.},
	author = {Long, Gary L. and Winefordner, J. D.},
	urldate = {2025-09-15},
	date = {1983-06-01},
	langid = {english},
}

@online{noauthor_iupac_nodate,
	title = {{IUPAC} - limit of detection (L03540)},
	url = {https://goldbook.iupac.org/terms/view/L03540},
	titleaddon = {{IUPAC} Gold Book},
	urldate = {2025-09-15},
}

@book{draper_applied_1998,
	location = {New York Chichester Weinheim [etc.]},
	edition = {3rd ed},
	title = {Applied regression analysis},
	isbn = {978-0-471-17082-2},
	url = {https://onlinelibrary.wiley.com/doi/book/10.1002/9781118625590},
	series = {Wiley series in probability and statistics},
	publisher = {J. Wiley \& sons},
	author = {Draper, Norman Richard and Smith, H.},
	date = {1998},
}

@book{massart_handbook_1997,
	location = {Amsterdam New York},
	title = {Handbook of chemometrics and qualimetrics},
	isbn = {0-444-89724-0},
	url = {https://pubs.acs.org/doi/10.1021/ci980427d},
	series = {Data handling in science and technology},
	number = {v. 20},
	publisher = {Elsevier},
	author = {Massart, Desiré Luc},
	date = {1997},
}

@article{rolo_hydrogen-based_2023,
	title = {Hydrogen-Based Energy Systems: Current Technology Development Status, Opportunities and Challenges},
	volume = {17},
	rights = {https://creativecommons.org/licenses/by/4.0/},
	issn = {1996-1073},
	url = {https://www.mdpi.com/1996-1073/17/1/180},
	doi = {10.3390/en17010180},
	shorttitle = {Hydrogen-Based Energy Systems},
	pages = {180},
	number = {1},
	journaltitle = {Energies},
	shortjournal = {Energies},
	author = {Rolo, Inês and Costa, Vítor A. F. and Brito, Francisco P.},
	urldate = {2025-09-16},
	date = {2023-12-28},
	langid = {english},
}

@article{farias_use_2022,
	title = {Use of Hydrogen as Fuel: A Trend of the 21st Century},
	volume = {15},
	rights = {https://creativecommons.org/licenses/by/4.0/},
	issn = {1996-1073},
	url = {https://www.mdpi.com/1996-1073/15/1/311},
	doi = {10.3390/en15010311},
	shorttitle = {Use of Hydrogen as Fuel},
	pages = {311},
	number = {1},
	journaltitle = {Energies},
	shortjournal = {Energies},
	author = {Farias, Charles Bronzo Barbosa and Barreiros, Robson Carmelo Santos and Da Silva, Milena Fernandes and Casazza, Alessandro Alberto and Converti, Attilio and Sarubbo, Leonie Asfora},
	urldate = {2025-09-16},
	date = {2022-01-03},
	langid = {english},
}

@article{monks_atmospheric_2009,
	title = {Atmospheric composition change – global and regional air quality},
	volume = {43},
	rights = {https://www.elsevier.com/tdm/userlicense/1.0/},
	issn = {13522310},
	url = {https://linkinghub.elsevier.com/retrieve/pii/S1352231009007109},
	doi = {10.1016/j.atmosenv.2009.08.021},
	pages = {5268--5350},
	number = {33},
	journaltitle = {Atmospheric Environment},
	shortjournal = {Atmospheric Environment},
	author = {Monks, P.S. and Granier, C. and Fuzzi, S. and Stohl, A. and Williams, M.L. and Akimoto, H. and Amann, M. and Baklanov, A. and Baltensperger, U. and Bey, I. and Blake, N. and Blake, R.S. and Carslaw, K. and Cooper, O.R. and Dentener, F. and Fowler, D. and Fragkou, E. and Frost, G.J. and Generoso, S. and Ginoux, P. and Grewe, V. and Guenther, A. and Hansson, H.C. and Henne, S. and Hjorth, J. and Hofzumahaus, A. and Huntrieser, H. and Isaksen, I.S.A. and Jenkin, M.E. and Kaiser, J. and Kanakidou, M. and Klimont, Z. and Kulmala, M. and Laj, P. and Lawrence, M.G. and Lee, J.D. and Liousse, C. and Maione, M. and {McFiggans}, G. and Metzger, A. and Mieville, A. and Moussiopoulos, N. and Orlando, J.J. and O'Dowd, C.D. and Palmer, P.I. and Parrish, D.D. and Petzold, A. and Platt, U. and Pöschl, U. and Prévôt, A.S.H. and Reeves, C.E. and Reimann, S. and Rudich, Y. and Sellegri, K. and Steinbrecher, R. and Simpson, D. and Ten Brink, H. and Theloke, J. and Van Der Werf, G.R. and Vautard, R. and Vestreng, V. and Vlachokostas, Ch. and Von Glasow, R.},
	urldate = {2025-09-16},
	date = {2009-10},
	langid = {english},
}

@article{singh_high-precision_2023,
	title = {High-Precision Trace Hydrogen Sensing by Multipass Raman Scattering},
	volume = {23},
	issn = {1424-8220},
	url = {https://www.mdpi.com/1424-8220/23/11/5171},
	doi = {10.3390/s23115171},
	pages = {5171},
	number = {11},
	journaltitle = {Sensors},
	shortjournal = {Sensors},
	author = {Singh, Jaspreet and Muller, Andreas},
	urldate = {2026-01-19},
	date = {2023-05-29},
	langid = {english},
}

@article{xiao_overtone_2024,
	title = {Overtone Excitation of Nitrogen Molecules via Stimulated Raman Pumping},
	volume = {15},
	rights = {https://doi.org/10.15223/policy-029},
	issn = {1948-7185, 1948-7185},
	url = {https://pubs.acs.org/doi/10.1021/acs.jpclett.4c02608},
	doi = {10.1021/acs.jpclett.4c02608},
	pages = {11510--11516},
	number = {46},
	journaltitle = {The Journal of Physical Chemistry Letters},
	shortjournal = {J. Phys. Chem. Lett.},
	author = {Xiao, Yue and Wen, Liping and Li, Zhichao and Han, Jie and Wu, Wenjie and Wang, Tao and Xie, Yurun and Yang, Tiangang},
	urldate = {2026-01-20},
	date = {2024-11-21},
	langid = {english},
}

@article{singh_ambient_2023,
	title = {Ambient Hydrocarbon Detection with an Ultra-Low-Loss Cavity Raman Analyzer},
	volume = {95},
	rights = {https://doi.org/10.15223/policy-029},
	issn = {0003-2700, 1520-6882},
	url = {https://pubs.acs.org/doi/10.1021/acs.analchem.2c04707},
	doi = {10.1021/acs.analchem.2c04707},
	pages = {3703--3711},
	number = {7},
	journaltitle = {Analytical Chemistry},
	shortjournal = {Anal. Chem.},
	author = {Singh, J. and Muller, A.},
	urldate = {2026-01-20},
	date = {2023-02-21},
	langid = {english},
}

@article{arachchige_raman_2024,
	title = {Raman scattering applied to human breath analysis},
	volume = {177},
	issn = {01659936},
	url = {https://linkinghub.elsevier.com/retrieve/pii/S0165993624002747},
	doi = {10.1016/j.trac.2024.117791},
	pages = {117791},
	journaltitle = {{TrAC} Trends in Analytical Chemistry},
	shortjournal = {{TrAC} Trends in Analytical Chemistry},
	author = {Arachchige, Charuka Muktha and Muller, Andreas},
	urldate = {2026-02-03},
	date = {2024-08},
	langid = {english},
}

@article{velez_spontaneous_2021,
	title = {Spontaneous Raman scattering at trace gas concentrations with a pressurized external multipass cavity},
	volume = {32},
	issn = {0957-0233, 1361-6501},
	url = {https://iopscience.iop.org/article/10.1088/1361-6501/abd11e},
	doi = {10.1088/1361-6501/abd11e},
	pages = {045501},
	number = {4},
	journaltitle = {Measurement Science and Technology},
	shortjournal = {Meas. Sci. Technol.},
	author = {Velez, Juan S Gomez and Muller, Andreas},
	urldate = {2026-02-03},
	date = {2021-04-01},
	langid = {english},
}

@article{lehotay_application_2002,
	title = {Application of gas chromatography in food analysis},
	volume = {21},
	rights = {https://www.elsevier.com/tdm/userlicense/1.0/},
	issn = {01659936},
	url = {https://linkinghub.elsevier.com/retrieve/pii/S0165993602008051},
	doi = {10.1016/S0165-9936(02)00805-1},
	pages = {686--697},
	number = {9},
	journaltitle = {{TrAC} Trends in Analytical Chemistry},
	shortjournal = {{TrAC} Trends in Analytical Chemistry},
	author = {Lehotay, Steven J and Hajšlová, Jana},
	urldate = {2026-02-04},
	date = {2002-09},
	langid = {english},
}

@article{repa_analyses_2002,
	title = {Analyses of gas composition in vacuum systems by mass spectrometry},
	volume = {37},
	rights = {http://onlinelibrary.wiley.com/{termsAndConditions}\#vor},
	issn = {1076-5174, 1096-9888},
	url = {https://analyticalsciencejournals.onlinelibrary.wiley.com/doi/10.1002/jms.388},
	doi = {10.1002/jms.388},
	pages = {1287--1291},
	number = {12},
	journaltitle = {Journal of Mass Spectrometry},
	shortjournal = {J. Mass Spectrom.},
	author = {Řepa, P. and Tesař, J. and Gronych, T. and Peksa, L. and Wild, J.},
	urldate = {2026-02-04},
	date = {2002-12},
	langid = {english},
}

@article{jiao_outgassing_2019,
	title = {Outgassing Environment of Spacecraft: An Overview},
	volume = {611},
	issn = {1757-8981, 1757-899X},
	url = {https://iopscience.iop.org/article/10.1088/1757-899X/611/1/012071},
	doi = {10.1088/1757-899X/611/1/012071},
	shorttitle = {Outgassing Environment of Spacecraft},
	pages = {012071},
	number = {1},
	journaltitle = {{IOP} Conference Series: Materials Science and Engineering},
	shortjournal = {{IOP} Conf. Ser.: Mater. Sci. Eng.},
	author = {Jiao, Zilong and Jiang, Lixiang and Sun, Jipeng and Huang, Jianguo and Zhu, Yunfei},
	urldate = {2026-02-04},
	date = {2019-10-01},
	langid = {english},
}

@article{huang_-line_2017,
	title = {An in-line Mach-Zehnder Interferometer Using Thin-core Fiber for Ammonia Gas Sensing With High Sensitivity},
	volume = {7},
	issn = {2045-2322},
	url = {https://www.nature.com/articles/srep44994},
	doi = {10.1038/srep44994},
	pages = {44994},
	number = {1},
	journaltitle = {Scientific Reports},
	shortjournal = {Sci Rep},
	author = {Huang, Xinyue and Li, Xueming and Yang, Jianchun and Tao, Chuanyi and Guo, Xiaogang and Bao, Hebin and Yin, Yanjun and Chen, Huifei and Zhu, Yuhua},
	urldate = {2026-02-04},
	date = {2017-04-05},
	langid = {english},
}

@article{herzberg_possibility_1938,
	title = {On the Possibility of Detecting Molecular Hydrogen and Nitrogen in Planetary and Stellar Atmospheres by Their Rotation-Vibration Spectra},
	volume = {87},
	issn = {0004-637X, 1538-4357},
	url = {http://adsabs.harvard.edu/doi/10.1086/143935},
	doi = {10.1086/143935},
	pages = {428},
	journaltitle = {The Astrophysical Journal},
	shortjournal = {{ApJ}},
	author = {Herzberg, Gerhard},
	urldate = {2026-02-04},
	date = {1938-05},
	langid = {english},
}

@article{herzberg_quadrupole_1949,
	title = {Quadrupole Rotation-Vibration Spectrum of the Hydrogen Molecule},
	volume = {163},
	rights = {https://www.springernature.com/gp/researchers/text-and-data-mining},
	issn = {0028-0836, 1476-4687},
	url = {https://www.nature.com/articles/163170a0},
	doi = {10.1038/163170a0},
	pages = {170--170},
	number = {4135},
	journaltitle = {Nature},
	shortjournal = {Nature},
	author = {Herzberg, G.},
	urldate = {2026-02-04},
	date = {1949-01-29},
	langid = {english},
}

@article{zorin_advances_2022,
	title = {Advances in mid-infrared spectroscopy enabled by supercontinuum laser sources},
	volume = {30},
	issn = {1094-4087},
	url = {https://opg.optica.org/abstract.cfm?URI=oe-30-4-5222},
	doi = {10.1364/OE.447269},
	pages = {5222},
	number = {4},
	journaltitle = {Optics Express},
	shortjournal = {Opt. Express},
	author = {Zorin, Ivan and Gattinger, Paul and Ebner, Alexander and Brandstetter, Markus},
	urldate = {2026-02-04},
	date = {2022-02-14},
	langid = {english},
}

\end{document}